\documentclass[11pt,english,fleqn]{article}
\usepackage{graphicx,amsfonts,amssymb,amsmath,epsfig,amsthm,babel,multicol,fancyhdr, slashed,eucal}
\usepackage{simplewick} 
\usepackage[a4paper, total={16cm, 23cm}]{geometry}
\usepackage{appendix}
\usepackage{authblk}

\graphicspath{{home/}}

\numberwithin{equation}{section}

\newtheorem{definition}{Definition}

\newcommand{\intprod}{\boldsymbol{\mathbin{\raisebox{\depth}{\scalebox{1}[-1]{$\lnot$}}}}}
\newcommand{\bo}{\boldsymbol}
\newcommand{\smallfrac}[2]{{\textstyle\frac{#1}{#2}}}
\newcommand{\BE}{\begin{equation}}
\newcommand{\EE}{\end{equation}}
\newcommand{\mrm}{\mathrm}

\newcommand{\dd}{\mathrm{d}}

\newcommand{\me}{\mathrm{e}}
\newcommand{\mcal}{\mathcal}
\newcommand{\mn}{\mathnormal}
\newcommand{\del}{\partial}

\makeatletter
\newsavebox{\@brx}
\newcommand{\llangle}[1][]{\savebox{\@brx}{\(\m@th{#1\langle}\)}%
  \mathopen{\copy\@brx\kern-0.5\wd\@brx\usebox{\@brx}}}
\newcommand{\rrangle}[1][]{\savebox{\@brx}{\(\m@th{#1\rangle}\)}%
  \mathclose{\copy\@brx\kern-0.5\wd\@brx\usebox{\@brx}}}
\makeatother

\title{\textbf{Geometric Quantization}}
\author{Andrea Carosso\thanks{email: andrea.carosso@colorado.edu}}
\affil{\textit{\small{Department of Physics, University of Colorado, Boulder, Colorado 80309, United States}}}
\date{}

\begin{document}

\setlength{\abovedisplayskip}{4pt}
\setlength{\belowdisplayskip}{4pt}

\maketitle

\abstract

Geometric quantization is an attempt at using the differential-geometric ingredients of classical phase spaces regarded as symplectic manifolds in order to define a corresponding quantum theory. Generally, the process of geometric quantization is applicable to other symplectic manifolds, not only cotangent spaces. The resulting formalism provides a way of looking at quantum theory that is distinct from conventional approaches to the subject, e.g., the Dirac bra-ket formalism. In particular, such familiar features as the quantization of spin, the canonical quantization of position and momentum, and the  Schr\"{o}dinger equation all emerge from geometric quantization. This paper serves as a review of the subject written in an informal style, often taking an example-based approach to exposition, and attempts to present the material without assuming the reader is an expert in differential geometry.

\tableofcontents

\section{Introduction}

In the construction of a quantum theory from an associated classical theory, one employs a procedure called \emph{quantization.} This typically involves a mapping of classical observables to quantum operators that satisfies certain properties, such as Dirac's quantization conditions. For example, \emph{canonical} quantization maps the classical position $q$ and momentum $p$ of a particle in one dimension to the \emph{quantum operators} $\hat q$ and $\hat p$ that satisfy the rule $[\hat q, \hat p] = i\hbar \mathbb{I}$. The conceptual motivation for such a rule is to achieve the correct \emph{uncertainty relations}, which capture the statistical ambiguities inherent in the process of measurement of quantum systems. The quantum state of the system no longer consists of a trajectory in phase space, $(q(t),p(t))$, but rather a square-integrable, complex-valued \emph{wave function}, $\psi$, depending on the available position (\emph{or} momentum) coordinates of the associated classical particle, and which satisfies a \emph{Schr\"{o}dinger equation}, $i\hbar \del_t \psi = \hat h \psi$. The operator $\hat h$ is the operator corresponding to a classical Hamiltonian $h$, usually obtained by replacing the $q's$ and $p's$ by their corresponding operators according to some ordering prescription. \emph{Geometric quantization} (GQ) is the attempt to use the classical phase space of a physical system and the language of differential geometry in order to construct a quantum theory of that system, which satisfies a set of desired quantization conditions.

There is a representation of classical mechanics that emphasizes the differential-geometric\footnote{If the reader is unfamiliar with differential geometry, the first appendix serves as a basic introduction to the subject; a survey of phase spaces as symplectic manifolds will be given in section 2.} character of a system, namely, the representation as a symplectic manifold $(M,\omega)$, where $M = T^* Q$ is the cotangent bundle on the configuration space $Q$, with local coordinates $q$ on $Q$ and $p$ on the fibers. The bundle $T^* Q$ is called a \emph{phase space} in the physics literature. The symplectic structure $\omega$ is a differential 2-form, i.e., an antisymmetric two-index tensor, expressable in local coordinates by position and momentum as $\omega = \dd p_a \wedge \dd q^a$ with components $\omega_{ab}$.\footnote{Summation convention for the symplectic coordinates is used throughout this paper.} The dynamics of the system consists in the integral curves of a \emph{Hamiltonian vector field}, $X_h$, on the phase space $M$. This vector field, in turn, is determined by the specification of a Hamiltonian function $h$ on $M$. The set of vector fields $X_f$ for all functions $f$ on $M$ span an infinite dimensional Lie algebra, called the \emph{algebra of classical observables}. Symmetries of the system are those transformations (diffeomorphisms) on $M$ whose tangent vector fields commute (in the sense of Lie derivatives) with the Hamiltonian vector field. The hamiltonian vector fields generate \emph{canonical diffeomorphisms} on $M$.

The initial goal of GQ is to use the above features of a symplectic manifold to build a correct quantum mechanics (QM). To begin with, one observes that many of the important operators of a quantum theory are, in practice, differential or multiplicative operators on functions (wave functions) on phase space. Differential operators (of first order), however, constitute the vector fields on a  manifold, and we may regard any manifold as naturally coming with the space of functions defined on it. One therefore suspects that the geometric structures associated with a symplectic manifold might be sufficient for the construction of the states and operators of a corresponding quantum theory.

There are also quantum systems which are commonly said to have no classical counterpart. The most famous such system is that of a particle with \emph{spin}. The closest analog to such an object in classical mechanics is a spinning, rigid body, characterized by the constraint that the magnitude of its angular momentum about its center of mass is fixed, i.e., its value lives on a sphere. It will turn out that this system, too, defines a symplectic manifold (but not a cotangent bundle) whose quantization produces the usual quantum mechanical spin systems. Moreover, the relevant procedure may be generalized to Lie groups regarded as smooth manifolds whose symplectic structure is determined by coadjoint orbits \cite{Woodhouse1}, but this aspect will not be pursued in this paper, in the interest of pedagogy over generality.

The road to geometric quantization starts out quite promising; one finds the simple and pristine structure of prequantized manifolds, involving a quantization map that produces some of the familiar features of quantum mechanics: operators, Hilbert space, and quantum numbers. It quickly becomes apparent, however, that to get QM exactly right, prequantization is not enough. The Hilbert space is too large, many important operators are not correctly reproduced, and time evolution of states is not correct. A distinction is then drawn between \emph{prequantization} and \emph{quantization}. One therefore embarks on a quest to correct each of these issues until many of the usual results of QM are finally reproduced.  The resulting quantization procedure is not totally satisfactory, however; the framework becomes more and more abstract, and new problems arise just as the old problems are solved. The final product, although far from perfect, possesses an elegance of its own, and is good enough to describe many of the important quantum systems, such as Fock spaces in the holomorphic representation (including the harmonic oscillator), spin systems, and the position-space representation of quantum states which evolve according to the celebrated Schr\"{o}dinger equation.

It has been suggested (e.g., Blau \cite{Blau}) that the goal of obtaining quantum theory from classical systems is misguided. It is true that QM, being ``more fundamental,'' need not have any simple relationship with classical mechanics, other than reducing to it in some limit. One can argue that this is not quite the goal of GQ, however. First, if one regards classical mechanics as the mathematical description of physical systems based on specifying the time-evolution of point particles and the objects they comprise, then GQ certainly does more than try to obtain QM from classical systems. For, as already mentioned, the geometric approach seeks not just the quantization of cotangent spaces, but also of arbitrary symplectic manifolds.\footnote{We will find that some manifolds have many distinct quantum theories corresponding to them, as in the case of spin. Moreover, the choice of a polarization in the process of quantization will also spoil the notion of 1-1 correspondence between classical and quantum theories.} Second, the fact that Schr\"{o}dinger himself used classical Hamilton-Jacobi theory in a vitally important manner when deriving the wave equation, in the second 1926 paper on wave mechanics, suggests that QM is intimately connected with classical mechanics, in a way that conventional QM, as typically presented today, probably overlooks. Third, as Blau admits, the study of GQ may be worthwhile solely for the features of quantum theory which it reveals, even if the framework as a whole is misguided or impractical.

In the following, I provide an introductory account of GQ. I do not provide rigorous proofs for many of the claims that are made, but I do provide what is hopefully a sufficient justification for many of the claims. The presentation is a mixture of what is found in the wonderful book of Woodhouse \cite{Woodhouse1}, and the review articles of Blau \cite{Blau} and of Nair \cite{Nair}. Nair's article is perhaps closest to an introduction for physicists but suffers from notational deviations from the conventions of differential geoemetry, while Woodhouse's book is much more a text on mathematics, with numerous examples, but which occasionally fails to give thorough details of calculations that would benefit the presentation. I therefore have tried to fill in some of those gaps by working through several examples throughout the paper. Consequently, the framework of GQ in its full generality, e.g., the full formalism of metaplectic structures and coadjoint orbits, or the deep connections with representation theory, will not explicitly appear; this is a weakness of the paper.

\section{Classical systems} I first provide a rapid review of relevant structures from the symplectic geometry of classical mechanics. A classical system is well-described by a cotangent bundle, which is an instance of a symplectic manifold, but it is important to keep in mind that not all symplectic manifolds are cotangent bundles. Darboux's theorem, however, states that any symplectic manifold in some sense \emph{looks like} a cotangent bundle, if one restricts their attention to small enough (but finite) regions of the manifold.

\paragraph{Symplectic manifolds.}

Let $Q$ be a configuration space of dimension $n$. The cotangent bundle is obtained by associating a \emph{momentum space} to each point $q \in Q$, spanned by 1-forms $\dd q^a$. Because velocities $\dot q$ naturally belong to the tangent spaces on $Q$, the momenta $p$ naturally belong to the cotangent spaces. Elements of the cotangent bundle are then expressable by $p_a \dd q^a$. The bundle is denoted by $T^*Q$, and has the natural projection map $\pi: T^* Q\rightarrow Q$; the projection of a point $(q,p)$ is $\pi(q,p) = q$.\footnote{I abuse notation, when no confusion arises, by labeling points of $T^* Q$ by their coordinates. I give a brief overview of fiber bundles in the appendix.} The \emph{fiber} $\pi^{-1}(q)$ above a configuration point $q$ is just the momentum space at $q$. The bundle $T^* Q$ represents the space of positions and momenta available to the physical system, that is, the phase space. 

$T^* Q$ defines a $2n$-dimensional symplectic manifold $(M,\omega)$, with $M = T^* Q$, and symplectic 2-form $\omega$. By Darboux's theorem \cite{Takhtajan}, $\omega$ is given in local symplectic coordinates by
\BE
\omega =  \dd p_a \wedge \dd q^a, \quad a = 1,\dots, n.
\EE 
Classical observables are given by functions on phase space, $f \in C^\infty(M)$, with real values $f(q,p) \in \mathbb{R}$. To every smooth $f$ there corresponds a \emph{hamiltonian vector field} $X_f$ defined implicitly by $\omega$ and the gradient 1-form,\footnote{I use $X \intprod \eta$ and $\eta(X,\bullet)$ interchangably, when no confusion arises, for the linear action of a differential form on a vector field in its first argument (the interior product). Thus we may write $\dd f + X_f \intprod \omega = 0$.}
\BE
\dd f + \omega(X_f, \bullet) = 0.
\EE
In local coordinates, $X_f$ is given by
\BE
X_f = \frac{\del f}{\del p_a} \frac{\del}{\del q^a} - \frac{\del f}{\del q^a} \frac{\del}{\del p_a}.
\EE
In particular, the vector fields generated by $q^a$ and $p_a$ are
\BE
X_{q^a} = -\frac{\del}{\del p_a}, \quad \mrm{and} \quad X_{p_a} = \frac{\del}{\del q^a}.
\EE
The Poisson bracket $\{f,g\}$ of two functions $f$ and $g$ is defined by 
\BE
\{f,g\} := X_f[g] = \dd g(X_f) = \omega(X_f,X_g) = - \{g,f\}.
\EE 
A short calculation implies that $[X_f,X_g] = X_{\{f,g\}}$; therefore, the space of hamiltonian vector fields $V_H(M)$ is an (infinite-dimensional) Lie algebra. A preferred energy function, called a \emph{Hamiltonian} $h$, determines the trajectories of the system via the integral curves $\gamma:[0,1]\rightarrow M, \; t \mapsto \gamma(t)$, of $X_h$, such that
\BE
\dot \gamma(t) = (X_h \circ \gamma)(t).
\EE
An observable $f$ is conserved, $(\dd/ \dd t)(f \circ \gamma) = 0$, whenever $\{f,h\}=0$. Since $X_0 = 0$, we have $\mcal{L}_{X_h} X_f = 0$ for such observables, where $\mcal{L}_X$ is the Lie derivative along $X$. 

The 2-form $\omega$ is closed, meaning that its exterior derivative vanishes: $\dd \omega  = 0$. Thus, by Poincar\'{e}'s Lemma, it is locally exact, $\omega = \dd \theta$. $\theta$ is known as the \emph{symplectic potential}. In local coordinates, 
\BE
\theta = p_a \dd q^a.
\EE 
$\theta$ is ``gauge-dependent'', in the sense that adding $\dd u$ to $\theta$ produces a different potential, but does not change the 2-form, since $\dd \dd u = 0$. The \emph{lagrangian} associated with any observable $f$ is defined by
\BE
L_f := X_f \intprod \theta - f.
\EE
A \emph{lagrangian submanifold} of $M$ is a $\mn\Lambda \subset M$ on which $\omega = 0$, with half the manifold dimension, $\dim \mn\Lambda = \dim M /2$. It follows that locally, $\theta |_{\mn\Lambda} = \dd S|_{\mn\Lambda}$ is exact; the function $S$ is called the \emph{generator} of the submanifold, and the fiber coordinates $p_a$ on $\mn\Lambda$ are determined by
\BE
p_a = \frac{\del S}{\del q^a}.
\EE
This is a familiar formula from Hamilton-Jacobi theory, and when $S$ is identified with a generator of canonical transformations produced by the hamiltonian vector field $X_h$, one can identify $S$ as the prinicpal function.

\paragraph{Canonical diffeomorphisms.} A diffeomorphism $\rho: M \rightarrow M$ is \emph{canonical} when it preserves the symplectic form,  $\rho^*\omega = \omega$. The exponentiation of a hamiltonian vector field is a canonical diffeomorphism on $M$. In turn, the flow generated by a hamiltonian vector field $X_f$ is preserved under $\rho$, in the sense that the push-forward of the tangent $\dot \gamma$ to integral curves of $f$ by $\rho$ is also a hamiltonian flow, which is generated by the new function $k = f \circ \rho^{-1}$ \cite{Takhtajan}.

Given two configuration spaces $Q, \; Q'$, every smooth $S\in C^\infty(Q\times Q')$ determines a canonical transformation $\rho : T^*Q \rightarrow T^*Q'$. Conversely, almost every canonical transformation may be determined by a generating function $S$.\footnote{The exception is when the $\mn\Lambda$ associated with $S$ is not a submanifold of $T^*(Q\times Q')$ \cite{Woodhouse1}.} Given $S$, the canonical transformation $\rho$ is given by solving
\BE
p_a = \frac{\del S}{\del q^a}, \quad p'_a = \frac{\del S}{\del q'^a},
\EE
for $p', q'$, the final coordinates, as functions of $q,p$, the initial coordinates.

I have included a sketch of the proof that the generating function $S$ of canonical transformations $\rho_t$ induced by the hamiltonian $h$ is indeed given by the action functional, since the proof is elegant, and provides a different perspective from the usual derivations found in physics texts. Begin with the definition of the lagrangian $L$ corresponding to $h$,
\BE
L = X_h \intprod \theta - h
\EE
(which is just the Legendre transform formula). From Cartan's formula,
\BE
\mcal{L}_X \theta = X \intprod \dd \theta + \dd ( X \intprod \theta),
\EE
and the definition of a hamiltonian vector field $X_h$, one finds
\BE
\mcal{L}_{X_h} \theta = \dd L.
\EE
Next, since the diffeomorphism generated by $X_h$ is canonical, it preserves the symplectic structure $\omega = \rho^* \omega$. Since locally $\omega = \dd \theta$, the 1-form $\theta - \rho^* \theta$ is closed and thus locally exact, hence
\BE
\theta - \rho^*_t \theta = \dd A
\EE
for some function $A$ on $M = T^* Q$, where $t$ parameterizes the integral curves $\gamma$ of $X_h$. Next, note that 
\BE
\frac{\dd}{\dd t} \theta(X_h) = \mcal{L}_{X_h}[\theta(X_h)] = X_h \intprod \mcal{L}_{X_h} \theta = X_h \intprod \dd L = \frac{\dd L}{\dd t},
\EE
by the formula for $\mcal{L}_{X_h} \theta$ above, and noting $\mcal{L}_{X_h} X_h = 0$. Integrating both sides produces
\BE
(\theta(X_h) \circ \gamma)(s) |^t_0 = (L\circ \gamma)(s) |^t_0.
\EE
Meanwhile, noting that the pushforward of a vector field along itself is an identity transformation, and using the local exactness of $\theta - \rho^*_t \theta$, one finds
\BE
(\theta(X_h) \circ \gamma)(s) |^t_0 = (\rho_s^* \theta) ( \rho_{s*}^{-1} X_h)(m)|^t_0 = (\rho_s^* \theta)(X_h)(m)|^t_0 = - (\dd A(X_h) \circ \gamma)(s)|^t_0,
\EE
where $\gamma(0) = m$. It follows upon integration that
\BE
A\circ \rho_t = - \int_0^t (L\circ\gamma)(s) \; \dd s + C.
\EE
Assuming that $X_h$ is complete, $A$ becomes a smooth function on $M$, and therefore depends on the phase space coordinates, $A(m,t), \; m \in M$. To obtain the usual principal function $S(q,q',t)$ of Hamilton-Jacobi theory, one uses $p = \del S/\del q$ which gives $p$ in terms of $q'$, allowing us to write
\BE
S(q,q',t) = A(q,p(q'),t) = -\int_0^t (L \circ \gamma)(s) \; \dd s + C, \quad \mrm{where} \quad \gamma(t) = \rho_t(q,p) = (q',p').
\EE
The integral is along the integral curve $\gamma$ of $X_h$ starting at $m = (q,p)$. One may lastly note that any canonical transformation generated by an observable $f$ will likewise have a generating function $S_f$, except the parameter $t$ will no longer be interpreted as time.

\paragraph{Symmetries.} The \emph{symmetries} of a classical system are those transformations by elements of a group $G$, such that $h \circ g = h, \; \forall g \in G$, where $g$ act on points of $M$ by $m\mapsto g(m) \in M$. \emph{Continuous} symmetries of a classical system are those $g$ which belong to a Lie group $G$ with algebra $\mathfrak{g}$. Every 1-parameter family $g_{t} = \me^{tA}$, $A \in \mathfrak{g}$, induces a vector field $X_{A}$ on $M$ tangent to the curve $\gamma(t) = g_t(m)$ via 
\BE
X_{A} [f](m) := \frac{\dd}{\dd t} f(\me^{tA}(m))|_{t=0}.
\EE
Then, one might look for a function $\mu_A \in C^\infty(M)$ on $M$, called a \emph{moment}, associated with every $A\in\mathfrak{g}$, such that the hamiltonian vector $X_{\mu_A}$ is equal to $X_{A}$. For example, in the case of $G = SO(3)$ on $M = \mathbb{R}^3$, the $g$ are rotations of the coordinates, and the moment of a rotation about $\bo n$ (corresponding to the group element $\exp \phi \bo \ell$,  where $\bo \ell = \bo n \cdot \bo X \in \mathfrak{so}(3)$), is the familiar function $\mu_{\bo \ell} = \bo n \cdot (\bo q \times \bo p)$ (the $X^i$ are a basis of $\mathfrak{so}(3)$). See \cite{Moit} for many more examples. I do not emphasize symmetries throughout the rest of the paper, so I will end my discussion of them here.

\section{Prequantization}

The goal of quantization is to construct a map $\mathbb{Q}$ from a classical observable $f \in C^\infty(M)$ to an operator $\mathbb{Q}(f) = \hat f$ and to define a Hilbert space $\mcal{H}$ of possible quantum states associated with the classical phase space $M$. Dirac was one of the first to propose a formal procedure for obtaining such operators in the late 1920's. Motivated by the successes of mapping classical Poisson brackets to quantum commutators, he suggested that the quantization map should obey the conditions (\cite{Woodhouse1}, \cite{Kirillov}):
\begin{enumerate}
\item Linearity: $\mathbb{Q}$ is linear; $\mathbb{Q}(f+g) = \mathbb{Q}(f) + \mathbb{Q}(g)$,
\item Commutators: $[\mathbb{Q}(f), \mathbb{Q} (g)] = - i\hbar \mathbb{Q} (\{f,g\})$,
\item Constants: for $f$ constant on $M$, $\mathbb Q (f) = f\mathbb{I}$ on $\mcal{H}$,
\end{enumerate} 
where $\hbar$ is Planck's constant. The last condition is required in order for canonical commutators to hold, which, in turn, are required in order to derive uncertainty principles such as that of position and momentum. Importantly, we must have 
\BE
[\hat q^a, \hat p_b] = i\hbar \delta^a_b\mathbb{I},
\EE
where $\hat q^a = q^a$ and $\hat p_a = -i\hbar \del/\del q^a$ on position-space states. One frequently also finds another condition, (5) Completeness: if $\{f_a\},\; i = 1,...,n$ form a complete set of classical observables, then the operators $\{\hat f_a\}$ are also complete. (A set of classical observables $\{f_a\}$ is \emph{complete} when $\{f_a, g\} = 0 \;\forall a$ implies that $g = \mrm{constant.}$ Likewise, a set of operators is complete if $[\hat f_a, \hat g] = 0 \; \forall a$ implies $g \propto \mathbb{I}$.) This property is equivalent to the condition that the operators $\hat f_a$ furnish an irreducible representation of the algebra of the classical observables $f_a$. We note that, by the deep Groenewold-van Hove (GvH) theorem \cite{Gotay}, no such ``Dirac'' map is sufficient for the construction of irreducible representations; extra conditions must be imposed, and this will be the central failure of our first attempt at quantization. Hence, what is achieved in this section is called a \emph{prequantization} (PQ), rather than a full \emph{quantization}.

\subsection{Quantum operators}

A first attempt to define $\mathbb{Q}$ is to map functions to their hamiltonian vector fields, $f \mapsto \hat f = -i\hbar X_f$. So, for example, $p_a \mapsto - i\hbar \del/\del q^a$, and $q^a \mapsto \hat q^a = i\hbar \del/\del p_a$. Although these operators look promising, the map is not good enough, for several reasons. First, the position operator would vanish on functions depending only on position, $\psi(q)$, when instead we want $\hat q^a \psi(q) = q^a \psi(q)$. More generally, the particular action of the operators should depend on whether the wave function is expressed in position or momentum space, while these operators are fixed by the form of the hamiltonian vector field. Second, since $[X_f, X_g] = X_{\{f,g\}}$ and $\{q^a,p_b\} = - \delta^a_b$, we have that $[\hat q^a, \hat p_b] = \hbar^2 X_{\delta^a_{\; b}} = 0$. 

Since the identity on a function space is multiplication by 1, and $X_f = 0$ for $f=$ constant, it's clear that the action of $\hat f$ on a function $\psi$ must include some kind of addition. The first modification would be $\hat f = -i\hbar X_f + f$, but then the commutator is $[\hat f, \hat g] = (-i\hbar)^2 X_{\{f,g\}} - 2i\hbar\{f,g\}$. The canonical commutation would then be $ [\hat q^a, \hat p_b] = 2i\hbar \delta_b^a$. This is closer, but there is still an unwanted factor of 2 that cannot be defined-away by any constant multiple of $f$ in $\hat f$.

The modification that corrects this issue is to define
\BE
f \longmapsto \hat f = -i\hbar X_f - X_f \intprod \theta + f,
\EE
where $\theta = p \cdot \dd q$ is the symplectic potential on classical phase space. From the identity
\BE
X[\eta(Y)] - Y[\eta(X)] = \dd \eta(X,Y) + \eta([X,Y]) \quad \mrm{with} \quad \eta \in \Omega^1(M),
\EE
and using $\dd \theta = \omega, \; [X_f,X_g] = X_{\{f,g\}}$, one finds
\BE
[\hat f, \hat g] = -i\hbar\Big(-i\hbar X_{\{f,g\}} - X_{\{f,g\}} \intprod \theta + \{f,g\}\Big) = -i\hbar \; \hat h,
\EE
where $h = \{f,g\}$. Since $X_{f+g} = X_f + X_g$, the map is linear, and for constant functions $f = c$, we have $\hat c = c$ because $X_c = 0$. Thus, all of Dirac's conditions are met.

The presence of the symplectic potential raises a few questions, however. First, not every symplectic manifold has a globally exact 2-form $\omega = \dd \theta$, in which case the manifold would not be quantized ``uniformly'' (this is not necessarily a bad thing; see below). Second, the potential is only defined up to a closed 1-form $\dd \phi$, since $\theta' = \theta + \dd \phi$ produces the same $\omega$, i.e., $\theta$ depends on choice of gauge. But then $\hat f$ changes by an amount $-X_f \intprod \dd \phi$, so that the quantization would have an unnatural non-uniqueness. If, however, the wave functions $\psi$ transform by a (local) phase $\psi'  = \me^{i\phi/\hbar} \psi$ whenever the potential changes by $\dd \phi$, then the $\frac{i}{\hbar} X_f[\phi] \psi'$ brought down by $X_f$ can be made to cancel the change in $\theta$. In fact, $-i\hbar X_f - \theta(X_f)$ has exactly the kind of form one expects from a covariant derivative.

The kind of structure we need is therefore nothing but a fiber bundle over $M$ whose fibers are $U(1)$, with connection 1-form $\Theta = \theta / \hbar$, and whose sections, i.e., things that the operators act on, are locally represented by complex wave functions (see appendix B for more on line bundles). The wave functions should be vectors in a Hilbert space, so the bundle must be equipped with a Hermitian inner product $\langle \cdot, \cdot \rangle$. The natural choice is
\BE
\langle \psi, \chi \rangle := \int_M (\psi, \chi) \bo \varepsilon,
\EE
where $\bo \varepsilon$ is the Liouville measure of the symplectic manifold, up to a multiplicative factor:
\BE
\bo \varepsilon = \frac{1}{(2\pi \hbar)^{n}} \; \omega \wedge \dots \wedge \omega.
\EE 
The conjugate-linear form $(\psi, \chi)$ is $\bar \psi \chi$ for complex scalar wave functions, or more generally $\psi^\dagger \chi$, where $\dagger$ is the conjugate-transpose of the vector $\phi \in \mathbb{C}^n$. Since the fibers are one-dimensional, the structure is also known as a \emph{Hermitian line bundle}. Thus, we characterize the process of prequantization by the following \cite{Woodhouse1}.
\begin{definition}[Prequantization]
A symplectic manifold $(M,\omega)$ is \emph{prequantizable} when there exists a Hermitian line bundle $\pi : B \rightarrow M$ with connection $\nabla$  and Hermitian form $(\cdot,\cdot)$, whose curvature $\Omega$ is proportional to the symplectic 2-form, $\Omega = \omega / \hbar$. The quantization map $\mathbb{P}$ carries classical observables $f$ to operators $\hat f$ via
\BE
\mathbb{P} : f \mapsto \hat f = -i\hbar \nabla_{X_f} + f,
\EE
which act on complex-valued sections (wave functions) $s_\psi = \psi \mathfrak{u} : M \rightarrow B$ living in the Hilbert space $\mcal{H}_\mathbb{P}$ of square-integrable (under $\langle \cdot, \cdot \rangle$) functions on $M$. The Hermitian form acts fiber-wise on sections by defining its action on the unit section $\mathfrak{u}$ (see the appendix),
\BE
(\mathfrak{u}, \mathfrak{u}) := 1,
\EE
so that on arbitrary sections $s_\psi, \; s_\phi$, one has $(s_\psi, s_\phi) = \bar \psi \phi$.

\end{definition}

There are several reasons why we have merely a \emph{prequantization} ($\mathbb{P})$, and not a \emph{quantization} ($\mathbb{Q})$. The first is that the wave functions, so far, depend on both position and momentum. In practice, wave functions depend on one or the other, not both. In particular, the operators $\mathbb{P}_q, \; \mathbb{P}_p$ are not the correct operators (see the next section). This is a manifestation of the GvH theorem; we have obtained a reducible, not irreducible, representation of the Heisenberg algebra, even though Dirac's conditions are met. This issue will be addressed by a choice of so-called polarization, described in section 4. Other short-comings of prequantization will be addressed throughout the remainder of this section. Thus, we denote the prequantization map by $\mathbb{P}$ rather than $\mathbb{Q}$.

One may characterize PQ as the result of regarding $\theta$ in much the same way one regards the vector potential $A$ of electromagnetism in standard QM. There, the electromagnetic field strength $F$ is derivable from a potential $A$ via exterior differentiation, $F= \dd A$.\footnote{This is the differential-geometric way of writing $F_{\mu\nu} = \del_\mu A_\nu - \del_\nu A_\mu$.} It is the potential which enters explicitly into covariant derivatives, $D = \dd -iA$, of wave functions $\psi$ on configuration space $Q$. The potential $A$ in this setting is a connection on a $U(1)$ bundle, and $F$ is the curvature. In the case of GQ, one still has a $U(1)$ bundle, but the connection is the symplectic potential $\theta/\hbar$, the curvature is the 2-form $\omega/\hbar$, and the covariant derivative is $\dd - i \theta/\hbar$. The similarity of these two situations motivates the definition of \emph{charged} symplectic structures, which will be discussed briefly in the next section. In electromagnetism, the potential has its own dynamics, while in GQ, the potential is, in some sense, fixed; the curvature $\omega$ is a given, fixed object, once the symplectic base space $M$ is given. $\theta$ can be any potential for which $\omega = \dd \theta$. 

\subsection{Weil integrality}

The existence of the prequantum line bundle $\pi:B\rightarrow M$, given a symplectic manifold $(M,\omega)$, is not guaranteed. The condition for its existence is called \emph{Weil's integrality condition}. A necessary condition (C1) is that the symplectic form should satisfy
\BE
\int_\Sigma \omega \in 2\pi \hbar \; \mathbb{Z},
\EE
for every closed 2-surface $\Sigma \subset M$. C1 is also sufficient when $M$ is simply connected \cite{Woodhouse1}. When $M$ is not simply connected, a more convenient condition (C2) of sufficiency is that the class of $\omega/2\pi\hbar$ in $H^2(M,\mathbb{R})$ should be in the image of $H^2(M,\mathbb{Z})$, that is, the coefficients of the vector space of closed-but-not-exact $\omega$'s are integers. 

We can understand the origin of the Weil integrality condition following the first edition of Woodhouse \cite{Woodhouse2}. Suppose that the bundle $B\rightarrow M$ exists. The \emph{parallel transport} of a section $s = \psi \mathfrak{u}$ along a curve $\gamma:[0,1]\rightarrow M, \; t\mapsto \gamma(t)$, with $\gamma(0) = m$, $\gamma(1) = m'$, and with tangent vector field $v = \dd/\dd t$, is determined by
\BE
\frac{\dd \psi}{\dd t} = \frac{i}{\hbar} (v \intprod \theta) \psi.
\EE
Together with an initial condition $(\psi\circ\gamma)(0) = \psi(m)$, integration of the ODE yields the unique result
\BE
\psi(t) = \exp\Big[\frac{i}{\hbar} \int_{\gamma_t} \theta \Big] \psi(0).
\EE
If $\gamma$ is a loop with $m=m'$, then Stokes' theorem implies
\BE
\oint_\gamma \theta = \int_{\Sigma_1} \dd \theta = \int_{\Sigma_1} \omega,
\EE
where $\Sigma_1$ is a 2-surface with boundary $\gamma = \del \Sigma_1$, and $\omega = \dd \theta$ is the symplectic 2-form. The parallel transport around $\gamma$ is then given by
\BE
\psi(t) = \exp\Big[\frac{i}{\hbar} \int_{\Sigma_1} \omega \Big] \psi(0).
\EE
Now imagine taking $\gamma$ as the boundary of a second surface $\Sigma_2$, such that $\Sigma = \Sigma_1 \cup \Sigma_2$ is a \emph{closed} 2-surface in $M$, that is ``cut'' into two halves by $\gamma$. Then it must be true that also
\BE
\psi(t) = \exp\Big[-\frac{i}{\hbar} \int_{\Sigma_2} \omega \Big] \psi(0),
\EE
the minus sign coming from the fact that $\del \Sigma_2 = - \gamma$ in order to get the orientation of $\Sigma_2$ right:
\BE
\oint_{\gamma} \theta = - \oint_{-\gamma} \!\!\!\theta = - \int_{\Sigma_2} \omega.
\EE
By the uniqueness of the ODE, one must obtain the same solution $\psi(1)$ at the final time $t=1$ from both surfaces $\Sigma_1, \; \Sigma_2$, so the phases must equal eachother:
\BE
\exp\Big[\frac{i}{\hbar} \int_{\Sigma_1} \omega \Big] = \exp\Big[-\frac{i}{\hbar} \int_{\Sigma_2} \omega \Big].
\EE
Bringing both factors to the same side yields the result
\BE
\exp\Big[\frac{i}{\hbar} \oint_{\Sigma} \omega \Big] = 1.
\EE
Thus, in order for the symplectic 2-form $\omega$ to be the curvature of a line bundle over $M$, it must satisfy Weil's integrality condition $\oint_\Sigma \omega = (2\pi\hbar)n\;$! It is this very integrality that leads to the quantization of spin in quantum theory, according to GQ. We will see several examples of prequantizations in the following few subsections.

\paragraph{Canonical quantization.} In the case where the symplectic manifold is a trivial phase space (cotangent bundle) $T^* Q \cong \mathbb{R}^{2n}$, with symplectic form $\omega = \dd p_a \wedge \dd q^a$, the Weil integrality condition is trivially satisfied, i.e., the integer is $n=0$. This is just Stokes' theorem on a simply-connected manifold:
\BE
\int_\Sigma \omega = \oint_{\del \Sigma} \theta = 0,
\EE 
since $\del \Sigma = 0$ for closed $\Sigma$, and $\omega = \dd (p\cdot \dd q)$ is globally exact. The operators corresponding to the position and momentum under the prequantization map $\mathbb{P}$ are
\BE
\hat q^a = i\hbar \frac{\del}{\del p_a} + q^a,  \quad \hat p_a = -i\hbar \frac{\del}{\del q^a}.
\EE
Since the sections of $B \rightarrow M$ depend, in general, on all the phase space coordinates $(p,q)$, we do not have the correct quantum operators. The prequantum Hilbert space $\mcal{H}_\mathbb{P}$ is the space of sections whose trivializations are square-integrable, $\psi \in L^2(\mathbb{R}^{2n})$ for $s = \psi \mathfrak{u}$. The inner product is given, in terms of local coordinates, by
\BE
\langle \psi, \chi \rangle = \int_{\mathbb{R}^{2n}} \!\! \frac{\dd^n q \dd^n p}{(2\pi\hbar)^n} \; \bar\psi(q,p) \chi(q,p).
\EE
From a group theory perspective, we see that the exponentiation of $\hat{p}_a$ gives the correct translation operator on wave functions, but the exponentiation of $\hat{q}^a$ does not give the correct momentum-space translation operator, due to the extra factor of $q^a$ which gets exponentiated. The problem is that we do not have an irreducible representation of the Heisenberg group. This is because the space of $p$-independent functions is closed under the action of $\hat q^a, \; \hat p_a$, which means the space of functions depending on \emph{both} $q$ \emph{and} $p$ is reducible. Full quantization will remedy this ailment. 

\paragraph{Spin quantization.} Prequantization does not require that the manifold $M$ is a cotangent bundle, only that it has a symplectic structure. A beautiful example is the case of the 2-sphere, $M = S^2$.

We can think of $S^2$ as the classical manifold of states of a spinning object with fixed magnitude $s$ of angular momentum, that is, a 2-sphere of radius $s$. $S^2$ is not the cotangent bundle of any configuration space, however.\footnote{One can regard $S^3 \cong SU(2)$ as a $U(1)$-bundle over $S^2$, and the quantization of the 2-sphere as a special case of the more general quantization scheme of $G/H$ spaces that produces realizations of irreducible representations of $G$. I will not pursue this characterization here.} The symplectic structure of $S^2$ is given by the volume-form $\omega \in \Omega^2(S^2)$ (typically called $\dd A$, the area element, in physics). In spherical polar coordinates $(\theta,\phi)$, the 2-form is
\BE
\omega =  s^2 \sin \theta \; \dd \theta \wedge \dd \phi.
\EE
The integral of $\omega$ over a closed 2-surface on the sphere is the well-known area $4\pi s^2$. For the prequantization $B\rightarrow M$ to exist, the integral of the curvature $\Omega = \omega / \hbar$ must  be an integer multiple of $2\pi$. Thus, there is a family of prequantizations of $S^2$ with symplectic 2-forms
\BE
\omega_{(n)} = n\hbar \omega, \quad n \in \mathbb{Z}.
\EE
The prequantum Hilbert space is then the space of square-integrable functions on the sphere, $L^2(S^2)$, an infinite dimensional Hilbert space. We will see in the section on holomorphic quantization that once we restrict this Hilbert space to that of holomorphic functions on $S^2$, we obtain exactly the irreducible representation spaces of $SU(2)$, i.e., the familiar spinor wave functions. If we rescale our symplectic form by dividing by $s$ to obtain
\BE
\omega := s \sin\theta \dd \theta \wedge \dd \phi,
\EE
then the resulting symplectic manifold is quantizable so long as 
\BE
s = \frac{\hbar}{2} \; n,
\EE
which is the familiar quantization of spin in QM. The rescaling by $s$ is less arbitrary from the point of view of quantization via coadjoint orbits \cite{Woodhouse1}. Further confirmation that $n$ indeed corresponds to spin will emerge once we introduce full quantization in section 4. It is quite remarkable that the quantization of $S^2$ so quickly results in the quantized aspect of spin. The fact that the same prescription leads to an almost-correct canonical quantization \emph{and} the quantization of spin is an inspiring feature that suggests GQ is on the right track.

\paragraph{Topological aspects.} If $M$ has a nontrivial cohomology group $H^1(M,\mathbb{R}) \neq 0$, then there exist closed 1-forms $A$ which are not exact. Manifolds with holes, like the torus, the circle, and a punctured plane, are examples of this. Such 1-forms have non-vanishing loop integrals when the loops enclose a hole. For example \cite{Blau}, on $S^1$ with angular coordinate $\phi$, the 1-form\footnote{There are quotes surrounding $\dd \phi$ because $\phi$ is not a function, since it is double-valued at $\phi = 0,\; 2\pi$; and the gradient $\dd$ is defined on \emph{functions}.}
\BE
A = ``\dd \phi" = \frac{x\dd y- y \dd x}{x^2 + y^2}
\EE
is closed, $\dd A = 0$. Naive application of Stokes' theorem would suggest that $\int_{\del D} A = \int_{D} \dd A = 0$, where $\del D = S^1$. But it is simple to show that the loop integral is in fact $2\pi n, \; n\in\mathbb{Z}$, where $n$ is the number of circuits around the circle executed by the loop, i.e., the winding number. Our naive Stokes argument was rendered invalid when we assumed that $A$ was defined on all of $S^1$; in fact, we need two ``patches'', or gauges, to integrate over the whole circle, owing to the two coordinate patches for $\phi$. Now consider the manifold $M = T^* S^1 \cong S^1 \times \mathbb{R}$. The (globally exact) symplectic form is
\BE
\omega = \dd p \wedge \dd \phi,
\EE
and a choice of symplectic potential is $\theta = p \dd \phi$. The prequantization of $M$ would seem to yield a line bundle $B\rightarrow M$, with connection $\nabla = \dd - i\theta/\hbar$. However, the closedness of $A$ means that we can add to $\theta$ any constant $\lambda$ times $A$ and obtain the same curvature $\omega$, implying a family of prequantizations $B_\lambda$, with associated potentials $\theta_\lambda$ and covariant derivatives $\nabla^{(\lambda)}$,
\BE
\theta_\lambda = \theta - \hbar \lambda A, \quad \nabla^{(\lambda)} = \dd - \frac{i}{\hbar} \theta + i\lambda A.
\EE
The prequantum operators corresponding to $p$ are
\BE
\mathbb{P}_p^{(\lambda)} = -i\hbar \del_\phi + \hbar \lambda.
\EE
The wave functions must be periodic in $\phi$, hence $\psi(\phi) \propto \me^{i n\phi}$ implies that the spectrum of $\mathbb{P}_p^{(\lambda)}$ is $\{(n + \lambda) \hbar, \; n\in\mathbb{Z} \}$. The range of $\lambda$ is $[0,1)$, as $\lambda$ and $\lambda + m, \; m\in\mathbb{Z},$ lead to the same spectrum of $\mathbb{P}_p^{(\lambda)}$. Thus, the family of prequantizations $B_\lambda$ are \emph{inequivalent}. Similarly, manifolds with nontrivial $H^2(M,\mathbb{R})$ will in general have inequivalent prequantizations; the quantization of $S^2$ was an example of this, in which the inequivalent prequantizations correspond to systems with different total spin.

From the path integral perspective, we can understand this example as a (non-electromagnetic) model of the Aharonov-Bohm effect. The propagation amplitude from a point $a$ to another one $b$ will be affected by the connectedness of the configuration space. If there is a hole in the plane $\mathbb{R}^2$, for example, we get the cohomology class $H^1(M,\mathbb{R}) = U(1)$ of $\mathbb{R}^2-\{0\}$, of which $A$ is the generator. In a sector where the symplectic potential is $\theta_\lambda$, the integration over paths will receive different phase contributions from paths which pass below and above the hole, leading to interference effects \cite{Baez}.

Dirac's electric charge quantization conjecture by the existence of magnetic monopoles also follows quickly from the topological aspects of prequantization. One may define a \emph{charged} symplectic structure on a cotangent bundle by $\omega_F = \omega + eF$, where $F\in\Omega^2(Q)$ is the electromagnetic field strength tensor. If $F$ is a non-trivial element of $H^2(M,\mathbb{R})$, as it is for the magnetic monopole \cite{Baez}, then the Weil integrality condition immediately implies that $e$ is quantized.

\subsection{Unitary evolution} Recall that classical canonical tranformations are those which preserve the symplectic structure, $\rho^* \omega = \omega$. Every hamiltonian vector $X_f$ generates a canonical flow $\rho_t$. In particular, the Hamiltonian $h$ generates classical time evolution, described by the integral curves of $X_h$, acting on points of phase space by mapping $m \mapsto \rho_t(m)$. This induces a time evolution of functions via pullback, $f_t (m) := (\rho_t^* f)(m) = (f \circ \rho_t)(m)$. $X_h$ generates the time evolution in the sense that
\BE
\frac{\dd}{\dd t} f_t(m) = \frac{\dd}{\dd t} f( \rho_t (m))= X_h[f](\rho_t (m)).
\EE 
Since $X_h[f] = \{ h, f\}$, this is equivalent to the more familiar formula $\dot f = \{h,f\}$. We will soon discover that the prequantum operators $\mathbb{P}_f$ similarly determine a time evolution of \emph{sections} of the bundle $\pi:B\rightarrow M$.

We begin by recalling how usual horizontal lifts induce parallel transport of sections. Let $X$ generate a flow $\gamma$ on $M$. The horizontal lift of $\gamma$ is a curve $\tilde \gamma$ in $B$ with tangent $\tilde X \in TB$ such that $\alpha(\tilde X) = 0$ (the \emph{horizontality condition}), where $\alpha = \pi^*\theta/\hbar + i\dd z/z$ is the connection 1-form on the \emph{bundle}, and $z = \me^{i\phi}$ is a fiber coordinate.\footnote{Recall that the local connection 1-form $\theta$ is the pullback by a section $s$ of the connection form $\alpha$ on the bundle, $\theta = s^* \alpha$. I omit the pullback $s^*$ from now on.} The horizontality condition on $\tilde X = X + \dot z \del_z$ implies an ODE for $z(t) = (z \circ \gamma)(t)$:
\BE
\dot z(t) = \frac{i}{\hbar} (X \intprod \theta)(t) z(t) \quad \Rightarrow \quad z(t) = z(0) \exp \Big[ \frac{i}{\hbar} \int_{\gamma_t} \theta \Big].
\EE
The parallel transport $\bar s$ of a section $s$ of the bundle $B$, i.e., one which satisfies $\nabla_{X} \bar s = 0$, is then given by $\bar s(\gamma(t)) = z(t) s(\gamma(0))$, as we found in the section on Weil integrality. If $\gamma(t) = \rho_t(m)$ is the integral curve of a hamiltonian vector field $X_f$, then, for $s_\psi = \psi \mathfrak{u}$ and $\gamma(0) = m$, the parallel transport amounts to a multiplicative action $\bar \rho_t$ on $\psi$,
\BE
(\bar\rho_t \psi)(m) := \bar \psi(\rho_t m) = \exp \Big[ \frac{i}{\hbar} \int_{\gamma_t} \theta \Big] \; \psi(m).
\EE
We will find that the flow generated by $\mathbb{P}_f$ involves a ``mixture'' of the pullback action $\rho_t^*$ and the parallel transport $\bar\rho_t$.

Suppose that the prequantum operators $\mathbb{P}_f$ generate a flow on sections, in the sense that
\BE
-i\hbar \dot s_t = \mathbb{P}_f s_t,
\EE
where $s_t$ is the time-evolved section. Recalling the definition of the lagrangian corresponding to an observable  $f$, $L_f = X_f \intprod \theta - f$, one can show that 
\BE
\mathbb{P}_f = -i\hbar X_f - L_f.
\EE
Although this is not a vector field, due to the scalar term $-L_f$, it is a simple exercise (compute $X_f[L_f]$) to observe that the exponentiation of $i \mathbb{P}_f/\hbar$ produces a transformation $\hat \rho_t$ given by
\BE
(\hat \rho_t \psi)(m) = \exp\Big[-\frac{i}{\hbar} \int_0^t L_f \; \dd \tau \Big] \; (\rho_t^* \psi)(m).
\EE
We can gain a more geometric understanding of this evolution by defining a ``lift'' $V_f \in TB$ of hamiltonian flows $X_f$ via
\BE
V_f := X_f + \frac{i L_f}{\hbar} z \del_z,
\EE
where $z = r\me^{i\phi}$ is a fiber coordinate. It follows that $\alpha(V_f) = f \circ \pi / \hbar$, which is analogous to the horizontality condition $\alpha(\tilde X_f) = 0$. Writing $V_f = X_f + \dot z \del_z$, the ODE which determines the integral curves of $V_f$ (along the fibers) is
\BE
\dot z(t) = \frac{i}{\hbar} \big(X_f \intprod \theta - f \big)(t) z(t) \quad \Rightarrow \quad z(t) = z(0) \exp \Big[ \frac{i}{\hbar} \int_0^t (X_f \intprod \theta - f) \; \dd \tau \Big],
\EE
while the integral curves on the base $M$ are the hamiltonian paths $\dot \gamma = X_f$. Thus, $V_f$ generates a diffeomorphism $\xi_t : B \rightarrow B$ on the \emph{bundle} given by
\BE
\xi_t(m,z_0) = \Big(\rho_t m, z_0 \exp\Big[\frac{i}{\hbar} \int_\gamma L_f \Big]\Big)
\EE
Since $s$ has values in $B$, the action of $\xi_t$ on a value $s(m)$ produces another section 
\BE
\xi_t[ s(m)] = s(m) \; \exp\Big[\frac{i}{\hbar} \int_\gamma L_f \Big]
\EE
above the point $\rho_t m$. Thus, we find that the transformation generated by $\mathbb{P}_f$ acting on a section $s$ is the new section $\hat\rho_t s$ determined by
\BE
\xi_t [ (\hat \rho_t s)(m)] := s(\rho_t m),
\EE
since it follows that
\BE
(\hat \rho_t s)(m) = s(\rho_t m) \; \exp\Big[-\frac{i}{\hbar} \int_\gamma L_f \Big].
\EE
as we found earler. Thus, $\hat\rho_t s$ is the section that maps to $\rho_t^* s$ under the flow $\xi_t$ of $V_f$.

In the case of hamiltonian flow by the Hamiltonian, $f=h$, the lagrangian is the ``true'' lagrangian $L = X_h \intprod \theta - h$, i.e., the Legendre transform of $h$, so that the phase factor involves the classical action $S$:
\BE
(\hat \rho_t \psi)(m) = \me^{-iS[\gamma]/\hbar} (\rho_t^* \psi)(m),
\EE
where $S[\gamma]$ is evaluated on the path traced out by $\rho_t m$. It is clear that this evolution is not identical to standard time evolution in QM, which is given by $\psi(t) = \me^{-i t \hat{h}/ \hbar}\psi(0)$, since $\hat h = \mathbb{P}_h$ is not the correct hamiltonian operator. It follows from the identity
\BE
\mathbb{P}\big(f(q,p)\big) = \frac{\del f}{\del q^a} \mathbb{P}(q^a) + \frac{\del f}{\del p_a} \mathbb{P}(p_a) + f - \frac{\del f}{\del q^a} q^a - \frac{\del f}{\del p_a} p_a ,
\EE
that the Hamiltonian $h= p^2 / 2m$ maps to
\BE
\hat h = -\frac{i\hbar}{m} p_a \frac{\del}{\del q^a} - \frac{p^2}{2m},
\EE
instead of the known Laplacian operator $\hat h = - (\hbar^2/2m) \nabla^2_q$. The failure of PQ to correctly produce second (or higher) order operators is a difficult problem to solve. For the Schr\"{o}dinger operator, the problem will be solved by introducing the BKS construction of quantum operators. Interestingly, the evolution $\hat\rho_t \psi$ does, however, resemble the integrand of the path integral $\psi_t = \int \me^{iS[\gamma_t]/\hbar} \psi_0 \dd \gamma_t \dd\gamma_0$; Woodhouse pursues the relationship of GQ with the path integral formulation of QM further.

The flow generated by $\mathbb{P}_f$ nonetheless has some nice properties. First, it is unitary, in the sense that
\BE
\langle \hat \rho_t \psi, \hat \rho_t \chi \rangle = \langle \psi, \chi \rangle,
\EE
since the phase factor simply cancels out, and because
\BE
\int_M (\hat \rho_t \psi, \hat \rho_t \chi) \bo \varepsilon = \int_M (\rho_t \psi, \rho_t \chi) \bo \varepsilon = \int_{\rho_t^{-1}M} (\psi, \chi) \rho_t^* \bo \varepsilon = \int_M (\psi, \chi) \bo\varepsilon.
\EE
The last step follows from the invariance of the Liouville measure under canonical flows, and assuming that $X_f$ is complete. This is equivalent, by Stone's theorem, to the self-adjointness of the prequantum operators:
\BE
\langle \psi, \hat f \chi \rangle = \int_M (\psi, [-i\hbar\nabla_{X_f} + f]\chi) \bo\varepsilon = \int_M ([-i\hbar\nabla_{X_f} + f] \psi, \chi) \bo\varepsilon = \langle \hat f \psi, \chi \rangle.
\EE
The square-integrability of the prequantum sections $\psi, \chi \in \mcal{H}_\mathbb{P}$ is crucial here; without it, the boundary term from integration by parts might not exist, obstructing self-adjointness. Although all complete hamiltonian vectors fields generate unitary evolution, not all canonical transformations do; further restrictions arise from the topological properties of $M$ \cite{Woodhouse1}.

\section{Quantization}

The central unresolved issues with prequantization are (1) the dependence of wave functions and inner products on the \emph{entirety} of the symplectic manifold $M$, that is, reducibility, when we know that wave functions $\psi$ should depend only on subsets consisting of half the total number of coordinates, e.g., position $\psi(q)$ or momentum $\psi(p)$, but not both at once $\psi(q,p)$, and (2) the failure to properly quantize second (or higher) order operators, including the free-particle and harmonic oscillator Hamiltonians. The next step in GQ is then to remedy these failures. Essential to the solution of (1) is the introduction of the notion of a polarization of a symplectic manifold, which, roughly speaking, divides the symplectic coordinates into two equally-sized groups. A solution to $(2)$, called the BKS construction, is provided by modifying the pullback evolution described earlier, and essentially depends on the notion of polarization. Along the way, we will discover a few other problems which necessitate a redefinition of quantum operators (the metaplectic correction) and the appropriate inner product on the quantum Hilbert space (half-form quantization). Although all these modifications might be regarded as stains on the elegance of prequantization, the resulting formalism will turn out to have a beauty of its own. In what follows, I introduce the mathematical concept of a polarization, and proceed to give several well-understood examples where GQ is successful: holomorphic quantization (including the harmonic oscillator and spin systems) and cotangent spaces. I then define the BKS construction and derive the free-particle flat-space Schr\"{o}dinger equation.

\subsection{Polarizations} A \emph{polarization} of a symplectic manifold is a foliation of the manifold by Lagrangian subspaces. That is, a sub-bundle $P \subset TM$ such that $[X,Y]|_m \in P_m$ for all $X,Y \in P_m \subset T_m M$, and each $P_m$ is lagrangian with $\omega|_{P_m} = 0$. Suppose we also have a prequantization of $M$, namely, a Hermitian line bundle $B\rightarrow M$ whose connection is the symplectic potential. Once we have a polarization $P$, we can consider the \emph{constant} functions $f$ along $P$, which satisfy $X[f] = X \intprod \dd f = 0$ whenever $X\in P$. Since wave functions are in fact sections of $B$, however, what we need are the \emph{covariantly} constant sections, satisfying $\nabla_X s= 0, \; \forall X \in P$. The problem then arises of whether the quantum operators $\mathbb{Q}_f$ preserve the polarization,\footnote{I write $\mathbb{Q}_f$ for $\mathbb{Q}(f)$ to prevent a profusion of parentheses from here on out.} in the sense that 
\BE
\nabla_X s = 0 \quad \Rightarrow \quad \nabla_X \mathbb{Q}_f s = 0.
\EE
Since $\mathbb{Q}_f$ is linear in $\nabla_{X_f}$, what is needed is the commutation $\nabla_X \nabla_{X_f} = \nabla_{X_f} \nabla_X$, but this only holds when the curvature $\omega$ and $\nabla_{[X,X_f]}$ vanish. Generally, $\nabla_X \nabla_{X_f} = \nabla_{X_f} \nabla_X + \nabla_{[X,X_f]}  + i\omega(X,X_f)/\hbar$. Thus, for $\mathbb{Q}_f$ to preserve the polarization, one needs
\BE
[X,X_f] \in P.
\EE

For example, when the symplectic manifold is a cotangent bundle $T^* Q$, we might choose wave functions depending only on position, in which case $X[\psi] = 0$ for all $X = X_a \del_{p_a}$. The span of the momentum basis vectors at a point $(p,q)$ is a lagrangian subspace, since $\omega(\del_{p_a}, \del_{p_b}) = 0$, and the collection of all such subspaces over $T^* Q$ is called the \emph{vertical polarization} $P$. The usual choice of symplectic potential, $\theta = p_a \dd q^a$, satisfies $X \intprod \theta = 0, \; X\in P$. Such a potential is said to be \emph{adapted} to $P$; an adapted potential is convenient because the covariant derivative reduces to a partial derivative. One can check that the polarization-preserving condition on a function $f$ is $\del^2 f /\del p_a \del p_b= 0$, i.e., $f$ is at most linear in $p_a$:
\BE
f(q,p) = g(q) + h^a(q) p_a.
\EE
Thus, the space of classical observables $f$ which preserve quantum states under the quantization map $\mathbb{Q}$ are of a rather limited kind; in particular, the Hamiltonian $h \propto p^2$ does not preserve the polarization! This means that under the canonical flow of $h$, the wave function no longer depends only on the position coordinates, which is not correct.

Another important case is when the manifold to be quantized is a K\"{a}hler manifold $(M,\mrm{g},J)$, with scalar $\mcal{K}$. As a complex manifold, the tangent bundle is split into the disjoint eigenspaces of the complex structure $J$: $TM = T^+ M \oplus T^- M$. The Levi-Civita connection on $M$ preserves these subspaces. The symplectic potential and the K\"{a}hler 2-form are related to the scalar $\mcal{K}$ via
\BE
\theta = -i\del \mcal{K}, \quad \omega = i \del \bar \del \mcal{K}.
\EE
Prequantization determines a line bundle over $M$ viewed as a symplectic manifold with symplectic 2-form $\omega$, and with the associated Hilbert space of square-integrable complex functions.
Covariant derivatives then take the form
\BE
\nabla_X = X \intprod \big(\dd -\frac{1}{\hbar} \del \mcal{K} \big).
\EE
A polarization $P$ of $M$ then consists in a submanifold spanned by vectors $X$, say, on which $\omega$ vanishes. Wave functions are then chosen to be the sections polarized along $P$, that is,
\BE
\nabla_X s = 0 \; \quad \forall X \in P.
\EE
Furthermore, it is always possible to choose a gauge in which $\phi' = \phi \me^{-\mcal{K}/2\hbar}$ for holomorphic $\phi$ \cite{Woodhouse1}, so that the natural inner product $\langle \cdot, \cdot \rangle$ takes the form
\BE
\langle s, s'\rangle = \int_M (\phi, \phi') \me^{-\mcal{K}/\hbar} \bo \varepsilon.
\EE

When the observable $f$ does not preserve the chosen polarization, canonical evolution as defined in PQ de-polarizes the sections. This motivates a more general definition of quantum operators, which will be described in the section on time evolution.

Now that we have the apparatus of polarizations, we may define the quantization of a symplectic manifold.
\begin{definition}[Quantization]
A \emph{quantization} of a symplectic manifold $(M,\omega)$ is the prequantization $(B, \omega_{(n)})$ of $M$ together with a choice of polarization $P\subset TM$ of $M$. The Hilbert space $\mcal{H}_P$ of polarized, square-integrable sections of $B$ contains the \emph{$P$-wave functions} of the quantum system.
\end{definition}

The inner product on the space $\mcal{H}_P$ has not been specified, for as we shall soon see, it will generally depend on what kind of manifold $M$ is. In the K\"{a}hler case, the PQ inner product suffices, but for the cotangent bundle, for example, a new definition must be given.

\subsection{Holomorphic quantization} One of the great achievements of GQ is the quantization of K\"{a}hler manifolds, which reproduces the well-known Bargmann, or \emph{holomorphic representation}, of quantum mechanics. The procedure which achieves this is called \emph{holomorphic quantization}. Holomorphic quantization accomplishes the construction of bosonic Fock spaces in a very natural way. Applied to the harmonic oscillator, one obtains the correct holomorphic representation of the Hamiltonian along with the associated raising and lowering operators. One can also apply the formalism to $M=S^2$, which leads to the holomorphic representation of spin-$n/2$ systems.

\paragraph{Boson Fock spaces.} Take $M$ to be a $2n$-dimensional flat K\"{a}hler manifold with canonical coordinates $\{p_a, q^a\}, \; a = 1,...,n$. We can write the symplectic form and potential in terms of a K\"{a}hler scalar $\mcal{K}$ in holomorphic coordinates as
\BE
\omega = \dd p_a\wedge \dd q^a = \frac{i}{2}\dd z^a \wedge \dd \bar z^a = -i \del \bar \del \mcal{K}, \quad \theta = -i \del \mcal{K} = -\frac{i}{2} \bar z^a \dd z^a, \quad \mrm{and} \quad \mcal{K} = \frac{1}{2}\bar z^a z^a.
\EE
We observe that $\theta$ is adapted to the polarization $P$ spanned by the antiholomorphic basis $\{\del_{\bar a} \}$, i.e., $\del_{\bar a} \intprod \theta = 0$. The covariant derivative along $P$ in the $\theta$ gauge is then $\nabla_X = X \intprod \dd$, so that polarized sections\footnote{$\mathfrak{u}$ is the unit section, and $\psi$ is a representative of the section. See appendix B.}  $s_\psi = \psi \mathfrak{u}$ are simply the holomorphic functions on $M$,
\BE
\nabla_{\bar a} s_\psi = \del_{\bar a} \intprod \dd \psi \; \mathfrak{u} = 0, \quad \mrm{or} \quad \frac{\del \psi}{\del \bar z^a} = 0,
\EE
which implies $\psi(z,\bar z) = \phi(z)$ is holomorphic. Now consider the gauge determined by the potential
\BE
\theta_0 = \frac{1}{2} (p_a \dd q^a - q^a \dd p_a) = \theta + \frac{i}{2} \dd \mcal{K}.
\EE
Recalling that under a change of gauge $\theta' = \theta + \dd u,$ $\psi' = \me^{iu/\hbar} \psi $, we find that polarized sections in the $\theta_0$ gauge have the form
\BE
\phi_0(z, \bar z) = \phi(z) \exp\Big[-\frac{z \cdot \bar z}{4\hbar} \Big].
\EE
We now have all the ingredients to define the quantum Hilbert space $\mcal{H}$. It is the space of polarized sections on $M$ with finite inner product\footnote{I am brushing under the rug an issue that the scalar product $z \cdot \bar z = z^a \bar z^a$ really involves the metric $\mrm{g}$,  and will therefore need to be positive in order for integrals to converge.}
\BE
\langle s_\phi, s_{\phi'} \rangle = \int_M (\phi, \phi') \exp\Big[-\frac{z \cdot \bar z}{2\hbar} \Big]  \bo \varepsilon \quad < \infty,
\EE
where $\bo \varepsilon = \omega^n/ (2\pi \hbar)^n$ is the natural measure on $M$. The space $\mcal{H}$ coincides with the usual holomorphic representation, as described in \cite{JZJ}, for example. Since $\phi(z)$ is holomorphic on $M$, we may expand in the basis of polynomials $\{1,z^a, z^a z^b, ...\}$:
\BE
\phi(z) = \phi_0 + \phi_a z^a + \frac{1}{2!} \phi_{ab} z^a z^b + \cdots.
\EE
The subspace $S\mcal{H}$ of $\mcal{H}$ spanned by $\phi$ with symmetric tensor coefficients $\phi_{a_1 ... a_k} \; \forall k$ is the usual bosonic many-particle Fock space $S\mcal{H} = \mathbb{C} \oplus S\mcal{H}_1 \oplus S\mcal{H}_2 \oplus \dots$, and such $\phi$ are generating functions of the state vectors of the system. 

The Hamiltonian vector field of an $f$ (with the $\{ \del_a, \del_{\bar a} \}$ conventions in the appendix) is 
\BE
X_f = 2i\big( \del_{\bar z} f \cdot \del_z - \del_z f \cdot \del_{\bar z} \big).
\EE
Important examples are $f = z^a, \; \bar z^a,$ and $z \cdot \bar z$, for which
\BE
X_{z^a} = -2i\del_{\bar a}, \quad X_{\bar z^a} = 2i \del_a, \quad X_{z \cdot \bar z} = 2i( z\cdot \del_z - \bar z \cdot \del_{\bar z}).
\EE
The condition that a real observable $f(z,\bar z)$ preserves $P$ is then $[X_f, \del_{\bar a}] \in P$, or
\BE
\del_{\bar a} \del_b f = c_{a b}, \quad \mrm{and} \quad \del_a \del_b f = 0.
\EE
The first equation implies that $f$ contains $c_{ab} z^a \bar z^b$, the second equation implies that $f$ is quadratic in neither $z$ nor $\bar z$ (by taking the conjugate equation). Linear terms are allowed. Imposing $f \in \mathbb{R}$ then gives the general form
\BE
f(z,\bar z) = f_0 + w_a z^a + \bar w_a \bar z^a + c_{ab} z^a \bar z^b, \quad f_0\in \mathbb{R}, \;  c_{ab} = \bar c_{ba}.
\EE
It follows that the quantizations of $z^a,\; \bar z^a$, and $z \cdot \bar z,$ acting on $s_\phi = \phi \me^{-\mcal{K}/2\hbar} \mathfrak{u}$ in the $\theta_0$ gauge, amount to a linear operation on the holomorphic function $\phi$ given by\footnote{This is an abuse of notation; the operators $\hat f = \mathbb{Q}(f)$ act on \emph{sections}, not on the functions $\phi$ alone.}
\BE
\hat z^a \phi= z^a \phi, \quad \hat{\bar z}^a = 2\hbar \; \del_a \phi, \quad (z\cdot \bar z)^\wedge = 2\hbar \; z \cdot \del_z \phi.
\EE
The first two operators are holomorphic raising and lowering operators, and the third is \emph{almost} the harmonic oscillator Hamiltonian operator.

\paragraph{Harmonic oscillator.} 

This is the $n = \dim M = 2$ case of the Fock space construction above, with Hamiltonian $h = (p^2 + q^2)/2 = z\bar z / 2$. The quantization of $h$ therefore preserves the anti-holomorphic polarization, and the corresponding operator is
\BE
\hat h = \mathbb{Q}_h = \hbar  \; z \; \del_z,
\EE
which is almost equal to the well-known holomorphic representation of the harmonic oscillator Hamiltonian, i.e., the realization of the algebraic creation and annihilation operators. The eigenfunctions of $\hat h$ are the monomials in $z$: $\psi_n = z^n$ with eigenvalues $\hbar n$.

We observe that the important additive factor of 1/2 is missing from the operator $\hat h$, which is a manifestion of the usual operator-ordering problem of quantum mechanics in the GQ formalism. This is part of the motivation for a further modification to the quantization procedure, called the \emph{metaplectic correction}. The recipe for this correction goes as follows. Let $P = \mrm{span}\{X_l\}, \; l = 1,...,n=\frac{1}{2} \dim M,$ be a polarization, and suppose $f \in C^\infty(M)$ preserves the polarization, $[X_f, X_k] = A^l_k X_l \in P$. Let $A = [A^l_k]$ be the matrix of the coefficients $A^l_k$. Then, one defines the quantization of the observable $f$ by
\BE
\mathbb{Q}_f := -i\hbar \nabla_{X_f} + f - \frac{i\hbar}{2} \mrm{tr}(A).
\EE
For the antiholomorphic polarization spanned by $\del_{\bar a}$, the quantization of the $n$-oscillator Hamiltonian then results in the correct operator
\BE
\hat h = \hbar \big( z \cdot \del_z + \frac{n}{2}\big).
\EE
The spectrum is the usual $\{ \hbar(m_a+1/2), \; m_a \geq 0\}$ in each coordinate $z^a$, with eigenfunctions given by monomials $(z^a)^{m_a}$. The operator $\hat z^a$ is the raising operator, and $\hat{\bar z}^a$ is the lowering operator. The addition of the metaplectic correction to geometric quantization may seem rather ad hoc, but it turns out that a seemingly separate problem, which we will encounter in the subsection on phase space quantization, will be solved in such a way as to simultaneously implement the metaplectic correction. 

\paragraph{Spin quantization.} It is convenient to regard the 2-sphere as a K\"{a}hler manifold, whose coordinates are the complexification $(z,\bar z)$ of stereographic coordinates $(X,Y)$ in the plane $\mathbb{R}^2$. We need two patches $U_N, \; U_S$ to cover the sphere on the north and south poles, respectively. The coordinates on $U_S$, for example, are related to the Cartesian coordinates $(x^1, x^2, x^3)$ on $\mathbb{R}^3$ by
\BE
z = \frac{x^1 + ix^2}{1 - x^3}, \quad \bar z = \frac{x^1 - ix^2}{1 - x^3}.
\EE
These are well-defined at the south pole $x^3 = -1$, but singular at the north pole $x^3 = +1$. The coordinates on $U_N$ are defined with an opposite sign in the denominator. One can show that the family of symplectic 2-forms, which allow for the prequantization of $M$, are given in holomorphic coordinates by
\BE
\omega_{(n)} = i\hbar n \frac{\dd z \wedge \dd \bar z}{(1+ \bar z z )^2}.
\EE
The K\"{a}hler scalar and symplectic potential $\theta = -i\del \mcal{K}$ are simply
\BE
\mcal{K} = n\hbar \log(1+\bar z z), \quad \theta = -in \frac{\bar z \dd z}{1 + \bar z z}.
\EE
Choosing the polarization $P$ spanned by $\del_{\bar z}$, the wave functions become the functions $\psi(z) \me^{-\mcal{K}/2\hbar}$ with $\psi(z)$ holomorphic. The inner product with volume form $\bo\varepsilon = \omega/(2\pi\hbar)$ is given by
\BE
\langle \psi, \psi' \rangle = \frac{i}{2\pi} \int_{S^2} \frac{\dd z \wedge \dd \bar z}{( 1+ \bar z z )^2} \; \bar \psi(z) \psi(z) \; \me^{-\mcal{K} / \hbar} = \frac{i}{2\pi} \int_{S^2} \frac{\dd z \wedge \dd \bar z}{(1+\bar z z )^{n+2}} \; \bar \psi(z) \psi(z).
\EE
In stereographic polar coordinates $(X,Y) \rightarrow (R,\Theta)$, one has $z = R \me^{i\Theta}, \; \bar z z  = R^2$, and the volume form is
\BE
\frac{i}{2} \frac{\dd z \wedge \dd \bar z}{(1+\bar z z)^2} =  \frac{R \dd R \wedge \dd \Theta}{(1+R^2)^2}.
\EE
The product of two monomials $z^m, \; z^{m'}$ is then
\begin{align}
\langle z^m, z^{m'} \rangle &=\frac{1}{\pi} \int_0^\infty \int_0^{2\pi} \frac{R \dd R \wedge \dd \Theta}{(1+R^2)^{2+n}} \; R^m R^{m'} \me^{i(m' - m)\Theta} \\
& = \delta_{mm'} \; \frac{\Gamma(1 + m) \Gamma(1+n - m)}{\Gamma(n+2)}.
\end{align}
The gamma function $\Gamma(z)$ is singular on the negative integers $\mathbb{Z}^-$ including $0$. It follows that in the prequantization sector $\omega_{(n)}$, the holomorphic functions cannot contain powers of $z$ higher than $n$ in order to be integrable. A basis for our Hilbert space $\mcal{H}_n$ is then $\{1,z,\dots, z^n\}$, which has dimension $n+1$.  This fact suggests that $n=2j$ where $j$ is the spin quantum number. Defining $f_m^{(n)}(z) = C_m z^m$ to be orthonormal, one finds
\BE
\langle f_m^{(n)}, f_{m'}^{(n)} \rangle = \delta_{mm'} \; \frac{|C_m|^2}{(n+1)} \frac{m!(n-m)!}{n!} \quad \Rightarrow \quad f_m^{(n)}(z) = \Big[\frac{(n+1) \; n!}{m!(n-m)!}\Big]^{1/2} \; z^m.
\EE
This construction reproduces the coherent state formalism of spin states (see \cite{Radcliffe}, for example). One could go on following Nair \cite{Nair}, using the functions
\BE
J_+ = -\frac{n\hbar z}{1+ \bar z z}, \quad J_- = -\frac{n\hbar \bar z}{1+ \bar z z}, \quad J_3 = -\frac{n\hbar}{2} \frac{1-\bar z z }{1 + \bar z z},
\EE
and demonstrate that the hamiltonian vectors fields 
\BE
X_f = \frac{i}{n\hbar} (1+\bar z z)^2 \big( \del_{\bar z} f \del_z - \del_z f \del_{\bar z}\big)
\EE
of $J_\pm,\; J_3$ are the $\mathfrak{su}(2)$ isometries on $S^2$, and that the quantum operators $\mathbb{P}_f$ acting on the holomorphic functions $f_m^{(n)}(z)$ reproduce the usual spin operators, which then implies $j = n/2$ is the spin quantum number. I will move on to the quantization of phase spaces, however, in the interest of demonstrating how GQ finally accomplishes canonical quantization and the derivation of the Schr\"{o}dinger equation.

\subsection{Phase spaces}

When $M = T^* Q$ is a cotangent bundle, there always exists a \emph{vertical polarization} $P$ tangent to surfaces of constant $q$. That is, for fixed $q$, the momenta $p$ coordinatize the fiber above $q$, and the leaf $P_m$ is the span of the $\del_{p_a}$. The wave functions are the polarized sections $s_\psi$ such that $\nabla_X s_\psi = 0 \; \forall X \in P$. We sometimes write $B_P$ for the subspacec of the prequantum Hilbert space of $P-$polarized sections. Write $s_\psi = \psi \mathfrak{u}$. In the gauge determined by the canonical potential $\theta = p \cdot \dd q$, the covariant derivative is a partial derivative along fibers, so the $P$-polarized sections satisfy
\BE
v_a \frac{\del \psi}{\del p_a} = 0 \quad \forall X = v^a \del / \del p_a \in P.
\EE
In particular, $\del \psi / \del p_a = 0$, so that $\psi = \psi(q)$ only. Note that an ``opposite'' polarization can similarly be chosen, on which $\psi = \psi(p)$. The associated Hilbert space contains the square integrable, $P$-polarized sections. We then encounter a problem with the inner product:
\BE
\langle \psi, \psi' \rangle = \int_M (\psi, \psi') \bo \varepsilon = \infty,
\EE
since the integration over momenta $p_a$ diverges for general (non-compact) fibers $T_q^* Q$. Thus, the prequantum Hilbert space $\mcal{H}_\mathbb{P}$ is not an adequate quantum Hilbert space. This problem generically occurs once a polarization is picked; the case of K\"{a}hler manifolds was an exception.

The way out of the infinite-product problem is paved by the \emph{half-form quantization} scheme, which solves both the volume problem \emph{and} the ground state energy problem we encountered for the harmonic oscillator. 

\paragraph{Half-form quantization.}

Half-form quantization is, roughly speaking, the absorption of a certain ``square-rooted volume form'' into wave functions, such that the square of the wave function produces a finite measure on the submanifold $Q$ of $M$ picked out by a polarization $P$, rather than the full Liouville measure on $M$. To begin with, define the \emph{determinant bundle} $\mrm{det}(Q) = \bigwedge^n T^*Q^\mathbb{C}$, i.e., the complexified $n$-fold wedge product of cotangent bundles. Since the sections of $T^* Q$ are 1-forms, the $n$-fold wedge product is an $n$-form, so the sections of $\mrm{det}(Q)$ are complex ``volume forms'' $\beta$. To define these forms on $M$, one uses the pullback of the projection $\pi : M \rightarrow Q$, but I will not be careful about this distinction. Since $P$ is spanned by momentum basis vectors $\del_{p_a}$, the forms $\beta$ in $\mrm{det}(Q)$ satisfy
\BE
X \intprod \beta = 0, \quad \mrm{and} \quad X \intprod \dd \beta = 0, \quad \forall X \in P.
\EE
Having in mind the fact that the canonical symplectic potential $\theta$ is typically chosen to be adapted to $P$, one defines the covariant derivative of a section $\beta$ of $\mrm{det}(Q)$ by $\nabla_X \beta = X \intprod \beta$.\footnote{This covariant derivative defines, in fact, a \emph{flat connection} with $[\nabla_X, \nabla_Y] = \nabla_{[X,Y]}$.} One also has, of course, the Lie derivative $\mcal{L}_X \beta$.

One then defines \emph{half-forms} $\nu$ such that $\nu^2 = \beta$. The $\nu$ are sections of a ``square-root'' bundle $\delta_P$, with transition functions given by the square roots of transition functions of $\mrm{det}(Q)$. The derivatives $\nabla$ and $\mcal{L}$ on such $\nu$ are defined by
\BE
\nabla_X \nu^2 = 2 \nu \nabla_X \nu, \quad \mrm{and} \quad \mcal{L}_X \nu^2 = 2\nu \mcal{L}_X \nu.
\EE
The \emph{quantum line bundle} corresponding to the polarized manifold $M$ is then obtained by the product of the polarized prequantum line bundle $B_P\rightarrow M$ with the square-root bundle $\delta_P$, so $B_P \otimes \delta_P \rightarrow M$ is the full bundle. Sections of this bundle have the form $\tilde s = s \nu$, where $s$ is a section of $B_P$. $P$-polarized wave functions are those for which $\nabla_X \tilde s = 0$, or
\BE
(\nabla_X s) \nu + s (\nabla_X \nu) = 0, \; \quad \forall X \in P.
\EE
The Hermitian structure is modified to include a product of half-forms,
\BE
(\tilde s, \tilde s') := (s, s') \bar \nu \nu ',
\EE
and their covariant derivative along $X \in P$ satisfy $\nabla_X (\tilde s, \tilde s') = 0$, since the Leibniz rule distributes $\nabla_X$ into each factor, and each one vanishes if $\tilde s, \; \tilde s'$ are both polarized. Thus, $(\tilde s, \tilde s')$ is an $n$-form on $Q$, and naturally defines the desired integration for the full inner product:
\BE
\langle \tilde s, \tilde s' \rangle := \int_Q (\tilde s, \tilde s').
\EE
Finally, quantum operators corresponding to $P$-polarization preserving classical observables $f = v(q)\cdot p + u(q) \in C^\infty (M)$ are defined on sections $\tilde s$ as
\BE \label{hfop}
\mathbb{Q}_f \tilde s := (\mathbb{P}_f s) \nu - i \hbar s \; \mcal{L}_{X_f} \nu.
\EE

We may now carry out the process of canonical quantization. Let $\tilde s = s\nu$, where $s = \psi \mathfrak{u}$ is a $P$-polarized section, i.e., $\psi = \psi(q)$. The polarization $P$ determines a volume form on $Q$ denoted by $\mu = \nu^2$. I will abbreviate such sections by $\tilde s = \psi \sqrt{\mu}$. The hamiltonian vector of $f$ is given by
\BE
X_f = v - (p_b \del_{q^a} v^b + \del_{q^a} u) \del_{p_a},
\EE
where $v = v^a(q) \del_{q^a}$ is a vector field on $Q$. In the $\theta = p \cdot \dd q$ gauge, the $v\cdot p$ term from $X_f \intprod \theta$ cancels the corresponding term in $f$, so the prequantization of $f$ is then
\BE
\mathbb{P}_f = -i \hbar v  + u.
\EE
The Lie derivative term in $\mathbb{Q}_f$ is evaluated by noting that $\del_p \intprod \mu = \del_p \intprod \dd \mu = 0$, so
\BE
\mcal{L}_{X_f} \mu^{1/2} = \frac{1}{2} \mu^{-1/2} \mcal{L}_{X_f} \mu = \frac{1}{2} \mu^{-1/2} \mcal{L}_v \mu = \frac{1}{2} \mrm{div}(v) \mu^{1/2}.
\EE
Plugging into the definition of $\mathbb{Q}_f$, the final result is
\BE
\mathbb{Q}_f (\psi \nu) = -i\hbar v[\psi] \nu + \big(u - \frac{i\hbar}{2} \mrm{div}(v)\big) \psi \nu.
\EE
The canonical quantum operators corresponding to $q^a$ and $p_a$ then amount to an action:
\BE
\hat q^a \psi = q^a \psi, \quad \mrm{and} \quad \hat p_a \psi = -i\hbar \frac{\del \psi}{\del q^a},
\EE
which are the canonical quantization relations at the heart of quantum theory. Moreover, they furnish an irreducible representation of the Heisenberg algebra, satisfying the canonical commutation relation. One can also demonstrate that these operators are self-adjoint:
\BE
\langle \tilde s, \hat{f} \tilde s' \rangle = \langle \hat f \tilde s, \tilde s' \rangle,
\EE
as desired. It is not difficult to show that this construction reproduces the metaplectic correction, since $[X_f, \del_p] = \del_{q^b} v^a \del_{p_b}$ and $\mrm{tr}(\del_{q^b} v^a) = \mrm{div}(v)$.○	

\subsection{Time evolution}

I will provide a brief account of how time evolution is ultimately defined for the polarized sections on a prequantum bundle in the case of the cotangent bundle, but will not present the full metaplectic formalism that is behind it. I reproduce some of the results of Woodhouse, while trying to provide more details of the actual computations involved in order to hopefully clarify the procedure, at the loss of generality.

Firstly, in the case where $P$ is preserved by an observable $f$, i.e., $[X_f,X]\in P \; \forall X \in P$, then the time evolution of a $P-$polarized section $s = \psi \mathfrak{u}$ of $B_P$ needs no further modification, for, as we saw in the discussion of unitary evolution in PQ, the operators $\mathbb{P}_f$ are generators of unitary time evolution:
\BE
-i\hbar \dot{s}_t = \frac{\dd}{\dd t'} (\hat\rho_{\delta t} s_t)|_{t'=t} = \mathbb{Q}_f s_t
\EE
where $\delta t = t' - t$ and $\mathbb{P}_f = \mathbb{Q}_f$ on such sections. If an initial section $\psi$ is polarized, then $\hat \rho_t \psi$ will also be polarized, since the operator commutation $[\nabla_X, \mathbb{P}_f] s = 0$ implies that $\nabla_X$ commutes with the exponentiation of $i\mathbb{P}_f /\hbar$. For the half-forms $\tilde s = s\nu$ in $B_P \times \delta_P$, one may use the pullback to define an evolution for $\nu$, leading to the full expression for $\tilde s_t = \tilde\rho_t \tilde s$,
\BE
\tilde \rho_t \tilde s := (\hat\rho_t s)(\rho_t^* \nu).
\EE
Taking the derivative at $t=0$, one reproduces the operator from Eq. (\ref{hfop}).

In the case of any higher-order differential operator, the evolution as previously defined will typically carry sections out of the polarization, so that the evolved sections will no longer be in the desired Hilbert space $\mcal{H}_P$. Yet, we know that certain quantum operators must be second-order differential operators. The definition of a quantum operator must therefore be modified, regardless of how elegant the prequantum operators are. The needed modification is a projection from the new Hilbert space $\mcal{H}_{P'}$ back onto the initial one. The structure which achieves this is a \emph{pairing}, and the definition of quantum operators that follows is called the \emph{Blattner-Kostant-Sternberg} construction. It (very nearly) produces the correct Schr\"{o}dinger operator, and is simply generalizable to Riemannian manifolds.

\paragraph{Pairing on cotangent bundles.} Given two Hilbert spaces $H_1, \; H_2$, a \emph{pairing} is a map $\llangle \cdot, \cdot \rrangle : H_1 \times H_2 \rightarrow \mathbb{C}$, which we can think of as a generalized inner product that ``mixes'' the two spaces. If $\langle \cdot,\cdot\rangle$ is the inner product on $H_1$, we may then define the \emph{projection} $\mn\Pi:H_2 \rightarrow H_1$ by
\BE
\langle v_1, \mn\Pi v_2 \rangle := \llangle v_1, v_2 \rrangle, \quad \mrm{where} \quad \mn\Pi v_2 \in H_1.
\EE
We will see this in action soon, where it reproduces the Fourier transform.

As an example, suppose that $M$ has two polarizations $P, \; P'$ which are transverse, so that $TM = P \oplus P'$. This implies that $M$ can be written as a product of two ``configuration spaces'' $Q, \; Q'$,  such that $M = T^* Q = Q\times Q'$; regarding $Q$ as the configuration space, then $Q'$ is the corresponding momentum space. Recalling that there almost always exists a generating function of canonical diffeomorphisms $S(q',q)$, the symplectic 2-form  can be written as
\BE
\omega = \del_{a} \del_{b'} S \;   \dd q'^b \wedge \dd q^a.
\EE
Next, one defines a pairing $(\cdot,\cdot)$ between the two bundles $\mrm{det}(Q), \; \mrm{det}(Q')$ by
\BE
(\beta, \beta') \bo\varepsilon := \beta \wedge \bar\beta',
\EE
recalling that $\beta, \; \beta'$ are complex-valued $n$-forms; so $(\beta, \beta') \in C^\infty(M)$, in contrast to the construction of the prior section on half-form quantization. The $\mrm{det}(Q)$ pairing induces a pairing on the half-form bundles $\delta_P, \; \delta_{P'}$ if one sets
\BE
(\nu , \nu') := \sqrt{(\nu^2, \nu'^2)} \in C^\infty(M),
\EE
in which $\nu^2,\; \nu'^2 \in \mrm{det}(Q)$. The pairing between Hilbert spaces corresponding to the polarizations $\mcal{H}_P, \; \mcal{H}_{P'}$ is then given by
\BE
\llangle \tilde s, \tilde s' \rrangle := \int_M (s,s')(\nu,\nu') \bo\varepsilon.
\EE
were $\tilde s = s\nu, \; \tilde s' = s' \nu'$ are sections of $B_P\times\delta_P, \; B_{P'} \times \delta_{P'}$ respectively. We can then define the projection of a section of $\mcal{H}_{P'}$ onto $\mcal{H}_P$ by
\BE
\langle \tilde s, \mn\Pi \tilde s' \rangle := \llangle \tilde s, \tilde s' \rrangle.
\EE

If we choose a gauge in which the potential is $\theta = \del_a S \dd q^a = \del S \cdot \dd q$, and let $P$ be the vertical polarization spanned by $\del_{p_a}$, then the $P$-polarized sections $s_\psi$ of $B_P$ having trivialization $\psi(q,q') = \phi(q)$ are independent of $p$. In the same gauge determined by $\theta$, however, the $P'$-polarized sections $s_{\psi'}$ satisfy (with a certain degree of sloppiness of notation)\footnote{One must bear in mind that sections polarized along distinct submanifolds generally look different in the same $\theta$ gauge---choice of polarization is distinct from choice of gauge.}
\BE
\nabla_{\del_{p'_a}} \psi' = \del_{p'_a} \psi' - \frac{i}{\hbar} \del_{p'_a} \intprod (\del S \cdot \dd q) = \del_{p'_a} \psi' - \frac{i}{\hbar} \del_{q^a} S \; \psi'= 0,
\EE
since, in this case, the $p'$ are identical to the $q$'s. This is because, assuming the polarizations are transverse, the surfaces of constant $q$ are the leaves of $P$, which must correspond to the ``base space'' of the polarization $P'$, whose leaves are surfaces of constant $q' = p$, and whose leaf coordinates are $p' = q$. Letting $\psi'(q',q) = \phi'(q') \exp R(q',p')$ one has (up to a constant)
\BE
\del_{p'_a} R = \frac{i}{\hbar} \del_{q^a} S \quad \Rightarrow \quad \psi'(q',q) = \phi'(q') \exp\Big[\frac{i}{\hbar} S(q',q)\Big].
\EE
The Liouville measure becomes, setting $\det{\del \del' S} = D$,
\BE
\bo\varepsilon = \frac{\omega^n}{(2\pi\hbar)^n} = \frac{D}{(2\pi \hbar)^n} (\dd q'^a \wedge \dd q^a)^n.
\EE
Now, choosing the trivialization $\nu = \sqrt{\dd^n q}, \; \nu' = \sqrt{\dd^n q'}$, we compute
\BE
(\nu^2,\nu'^2) \bo\varepsilon = (\nu^2,\nu'^2) \frac{D}{(2\pi \hbar)^n} \dd^n q \wedge \dd^n q' := \nu^2 \wedge \bar \nu'^2 = \dd^n q \wedge \dd^n q',
\EE
thus $(\nu^2, \nu'^2) = (2\pi\hbar)^n / D$, so that
\BE
(\nu, \nu') \bo\varepsilon = \frac{\sqrt{D}}{(2\pi\hbar)^{n/2}} \dd^n q \wedge \dd^n q'.
\EE
Setting $q' = p$ and noting that $S = p\cdot q$ (with trivial determinant $D$) generates the relevant canonical transformation, the pairing becomes
\BE
\llangle \tilde s, \tilde s' \rrangle = \frac{1}{(2\pi\hbar)^{n/2}} \int_Q \bar\phi(q) \Big[ \int_{Q'} \phi'(p) \exp\big[i p\cdot q/\hbar] \dd^n p \Big] \dd^n q.
\EE
The projection, finally, is therefore given by the Fourier tranform
\BE
(\mn\Pi \tilde s')(q) = \frac{1}{(2\pi\hbar)^{n/2}}\int_{Q'} \phi'(p) \exp\big[i p\cdot q/\hbar] \dd^n p \quad \in \mcal{H}_P,
\EE
which recovers the usual equivalence of representations between position and momentum space wave functions in QM. In GQ, therefore, the correct relationship between position and momentum space wave functions is regarded as an example of a canonical transformation achieved by use of a classical generating function $S= p\cdot q$; this is quite remarkable.

\paragraph{Blattner-Kostant-Sternberg construction.} Suppose we evolve a $P$-polarized wave function $\tilde s = s \nu$ along a hamiltonian vector field $X_f$, resulting in the flow $\tilde \rho_t \tilde s = (\hat \rho_t s)(\rho^*_t \nu)$, where $\hat\rho_t$ is the time-evolution described under prequantization. In general, $\tilde\rho_t \tilde s$ belongs to a new pulled-back polarization $P' = \rho_t^* P$ different from $P$. Since both Hilbert spaces must be subsets of the square-integrable sections on the prequantum bundle $B$, there should exist a pairing and a projection from one polarization to the other. This suggests the BKS time-evolution of a section $\tilde s_t$ at time $t$ defined by
\BE
\langle \dot{\tilde s}_t, \tilde r \rangle := -\frac{\dd}{\dd t'} \llangle \tilde \rho_{\delta t} \tilde s_t, \tilde r \rrangle|_{t'=0} = - \frac{\dd}{\dd t'} \langle \mn\Pi \tilde \rho_{\delta t} \tilde s_t, \tilde r \rangle|_{t'=0}, \quad \forall \tilde r \in \mcal{H}_P,
\EE
where $\delta t = t' - t$ and $\tilde r$ is independent of time. In other words, the generator of time evolution is determined by incrementing the sections forward along the prequantum flow of $X_h$ from time $t$ by a small amount $\delta t = t' - t$, then projecting the section back onto $\mcal{H}_P$, and finally taking a $\delta t \rightarrow 0$ limit. For simplicity, we will consider $\tilde s_t$ at $t=0$. The pairing on the RHS is then (for $\tilde r = \chi \sqrt{\mu}$)
\BE
\llangle \tilde \rho_t \tilde s, \tilde r \rrangle = \int_M \overline{(\rho_t^* \psi)} \; \chi \; \exp\Big[\frac{i}{\hbar} \int_0^t (L \circ \gamma)(t') \; \dd t' \Big] \sqrt{(\rho_t^* \mu , \mu)} \;  \bo \varepsilon.
\EE

I now present a derivation that the above definition implies the free-particle Schr\"{o}dinger equation in flat space $M = \mathbb{R}^{2n}$. 
It is a rather non-trivial calculation on a first approach, and the result is correct only up to a phase $c$.\footnote{Woodhouse performs the calculation for $h = \mrm{g}^{-1}(p,p)/2$, the free particle Hamiltonian on a Riemannian manifold with metric $\mrm{g}$.} To this end, let the Hamiltonian be $h = p^2/2m$. The Lagrangian is then $L = h$. The flow $\rho_t$ generated by $X_h$ is simply the inertial motion on flat space, that is, straight lines, with constant momenta:
\BE
q(t) = q + \frac{p}{m} t, \quad p(t) = p
\EE
(this is a vector equation, as $q \in \mathbb{R}^n$). The integral of $L$ over time is then simply $ht$. 

Next, we must compute the pullback of $\mu = \dd^n q$, the volume form on the initial manifold $Q$. We have to be rather careful here. Let $x : M \rightarrow \mathbb{R}^{2n}$ be the coordinate map from $M=T^* Q$ into $\mathbb{R}^{2n}$. Since $\rho_t : M \rightarrow M$ is a diffeomorphism, one has $(x\circ\rho_t)(m) = (q(t),p(t))$ as functions of the initial point $x(m)=(q,p)$. The pullback of a basis 1-form is
\BE
(\rho_t^* \dd x^\mu)(m) = \dd (x^\mu \circ \rho_t )(m) = \dd x^\mu (t), \quad \mu=1,...,2n.
\EE
Since $\dd^n q = \dd q^1 \wedge \cdots \wedge \dd q^n$, the pullback we need is found by computing
\BE
(\rho_t^* \mu, \mu) \bo\varepsilon := \rho_t^* \mu \wedge \mu = \dd^n(q +\frac{t}{m} p) \wedge \dd^n q = \Big(\frac{t}{m}\Big)^n \dd^n p \wedge \dd^n q.
\EE
It follows that the half-form contribution to the pairing is 
\BE
\sqrt{(\rho_t^* \mu, \mu)} = \Big(\frac{2\pi\hbar t}{m} \Big)^{n/2}.
\EE
The pairing is then
\BE
\llangle \tilde \rho_t \tilde s, \tilde r \rrangle = \int_M \overline{\psi(q(t),p(t))} \; \chi(q) \; \me^{ip^2 t/2m\hbar} \Big(\frac{t}{2\pi\hbar m} \Big)^{n/2} \dd^n p \wedge \dd^n q.
\EE
Since we want the time derivative of this quantity, we may simply find the $O(t)$ term in an expansion of the above formula. We also observe why the prequantum evolution carries $\tilde s$ out of $\mcal{H}_P$: $\psi(q(t),p(t))$ explicitly depends on the initial momenta $p$.

To obtain the projection $\mn\Pi \tilde \rho_t \tilde s$, we must evaluate the momentum integral
\BE
\int_{\mathbb{R}^n} \overline{\psi(q(t),p(t))} \; \me^{ip^2 t/2m\hbar}\dd^n p.
\EE
One approach is to rescale momenta by $t$ and use the saddle point approximation in the $t\rightarrow 0$ limit; this is discussed in Woodhouse. I follow a different route by use of the heat kernel. First, expand $\psi(q(t),p(t))$ about the initial point $q$. Noting that $p(t) = p$, and that $\rho^*_0 \psi$ is independent of $p$ (because it was $P$-polarized), 
\BE
\psi(q(t),p(t)) = \psi(q + tp/m, p) = \psi(q) + \frac{t}{m} \; p_a \del_a \psi(q) + \frac{t^2}{2m^2} p_a p_b \del_a \del_b \psi(q) + O(t^3),
\EE
The $O(p_a)$ integral vanishes by parity of the integrand. The $O(p_a p_b)$ term vanishes unless $a=b$, in which case the momentum integral is
\BE
\int_{\mathbb{R}^n} p_a^2 \me^{ip^2 t/2m\hbar}\dd^n p = \Big(\int_\mathbb{R} \dd p_a \; p_a^2 \me^{ip^2_a t/2m\hbar}\Big) \Big(\int_{\mathbb{R}^\ell} \me^{ik^2 t/2m\hbar}\dd^\ell k \Big),
\EE
where $\ell = n-1$. An explicit form for the second integral implies the result of the first integral upon differentiation and setting $\ell=1$. Setting $\tau = im\hbar/2t$, the second integral involves the well-known (analytically continued) infinite-space heat kernel
\BE
\int_{\mathbb{R}^\ell} \me^{-k^2/4\tau} \dd^\ell k = (4\pi \tau)^{\ell/2} \int_{\mathbb{R}^\ell} K_\tau(k) \dd^\ell k = (4\pi \tau)^{\ell/2}.
\EE
The entire momentum integral is therefore given by
\BE
\int_{\mathbb{R}^n} p_a^2 \me^{ip^2 t/2m\hbar}\dd^n p  = \frac{im\hbar}{t}\Big(\frac{2\pi i m \hbar}{t}\Big)^{n/2}.
\EE
Inserting this back into the pairing formula, we obtain
\BE
\llangle \tilde \rho_t \tilde s, \tilde r \rrangle = \llangle \tilde \rho_0 \tilde s, \tilde r \rrangle + i^{\frac{n}{2}} \frac{i\hbar t}{2m} \int_Q \nabla^2 \overline{\psi}(q) \; \chi(q) \dd^n q + O(t^2).
\EE
Taking the time derivative and evaluating at $t=0$, setting the result equal to $-\langle \dot{\tilde s}_0, \tilde r \rangle$, and taking the complex conjugate, one finds ($\forall \tilde r \in B \times \delta_P$)
\BE
i\hbar \frac{\del \psi}{\del t} = - \frac{c \hbar^2}{2m} \nabla^2 \psi.
\EE

It is remarkable that all of this abstract machinery has managed to finally produce the correct time evolution of wave functions; it is easy to lose hope along the way. The factor $c = \exp[-i\pi n/4]$ is an unwanted phase. One can do-away with it in the process of implementing the full-blown apparatus of metaplectic structures, which I have tried to avoid going into in this paper; the phase $c$ is then absorbed into a redefinition of the pairing. Thus, the failure of prequantization to correctly produce second-order operators has been solved by using the prequantum canonical transport $\tilde \rho_t$ of sections, followed by a projection $\mn\Pi$ back onto the initial polarization of the section. In general, the quantum operator $\mathbb{Q}_f$ corresponding to a classical observable $f\in C^\infty(M)$ gets the final definition for sections $\tilde s = s \nu$ in $\mcal{H}_P$,
\BE
\mathbb{Q}_f \tilde s := -i\hbar\frac{\dd}{\dd t} (\mn\Pi \tilde \rho_t \tilde s) |_{t=0}.
\EE
It is straight forward to check that this definition reproduces the previous definition of quantum operators on half-forms for polarization-preserving observables. Generalization to order three and higher operators is not simple, and here the formalism runs into problems: the operator definition above does not generally imply the commutation condition of Dirac; the operators are not always self-adjoint; the time evolution is not always unitary (although many important cases turn out to be so); there does not always exist a polarization on the symplectic manifold; the operators are not guaranteed to be linear \cite{Ali}. Thus, to obtain a correct quantum mechanics in GQ, one must be willing to abandon some of the initial postulates of quantization in their full generality.

An interesting aspect of the BKS procedure is the explicit role of classical trajectories, i.e., canonical flows generated by $X_h$. Part of the derivation requires expanding wave functions with flowed arguments about the initial points $(q,p)$ at $t$, and the precise form of this generally depends on the form of the canonical flows. This suggests that classical dynamics plays an important role in determining the evolution of wave functions. The symplectic potential together with the concept of canonical transformations also play an integral role; they are responsible for the interrelationship of position and momentum spaces, and they generate the action integral contribution to the phase of wave functions which is centrally important in the determination of the projection back to $\mcal{H}_P$, the polarized square-integrable sections. Even the Liouville measure remains significant, for it allows for the definition of a pairing of Hilbert spaces on the symplectic manifold. The introduction of half-forms is surprising at first, and one wonders how such an abstract object could be necessary in the construction of QM. They are important in providing the metaplectic correction, which fixes the vacuum energy of the harmonic oscillator, among other things; a rather mysterious connection. The relevance of such objects, however, is not \emph{as} surprising once one learns that half-forms can be thought of as a symplectic analog of spinors, which are ``square-roots'' of vectors on Riemannian manifolds; the half-forms, instead, are square-roots of volume forms.

\section{Summary}

Geometric quantization (GQ) is an attempt to better-define the quantization map first suggested by Dirac in such a way as to utilize the structure of classical symplectic manifolds. GQ occurs roughly in two steps: prequantization (PQ) and quantization. PQ consists of the construction of a prequantum line bundle over the symplectic manifold, a definition of quantum operators, and the identification a prequantum Hilbert space of square-integrable complex sections on the bundle. This construction correctly reproduces the three Dirac conditions for quantization. In particular, the commutation relation needed to obtain uncertainty relations is satisfied. In addition, the quantization of compact symplectic manifolds with non-trivial cohomology groups, like the 2-sphere, quickly leads to some characteristically ``quantum'' features, such as the quantization of spin.

PQ is not entirely satisfactory, however. The operator map does not correctly reproduce the familiar canonical quantization of position and momentum, and leads to the wrong free-particle hamiltonian operator, which implies an incorrect time evolution of wave functions. The solution to these failures begins with the introduction of polarizations of symplectic manifolds, which reduce the coordinate-dependence of sections to half that of the original manifold. In the K\"{a}hler case, this leads to the holomorphic formalism of quantum systems such as the harmonic oscillator, and spins systems. In the case of cotangent bundles, polarization leads to more familiar-looking wave functions, which depend on either position or momentum, but not both. The introduction of polarization brings a few new problems along with it, which motivates the redefinition of the wave functions of the system as including square-roots of volume forms on the classical configuration space. To properly define the time evolution of these half-forms, one needs to introduce the notion of a pairing of Hilbert spaces, which allows also for a method of relating wave functions from different polarizations; these modifications lead to the correct Fourier relationship between position and momentum space, and finally, to the correct Hamiltonian operator for wave functions in the position space representation.

In this paper, I have focused on presenting the basic aspects of GQ without too much discussion of the formalism in its full generality. One can further pursue GQ by generalizing these constructions. In particular, there is a general scheme for the quantization of Lie groups by use of a symplectic manifold consisting of coadjoint orbits of elements of the dual Lie algebra. This is the route pursued by some of the founders of the subject, Kostant, Sourieau, and Kirillov, and provides a link between the quantization of symplectic manifolds and the realization of irreducible representations of groups. One can also approach the problem by attempting to carry out quantization of arbitrary symplectic manifolds; since any symplectic manifold has local coordinates in which the full symplectic potential is determined by a mixture of a K\"{a}hler-like scalar with a canonical 1-form-like potential \cite{Woodhouse1}, this route is not too foreign to the constructions presented in this paper. I should note that the framework of GQ is also applicable to quantum field theory, for which the classical symplectic manifold is a cotangent space of an infinite dimensional configuration space of fields. Finally, the full implementation of the metaplectic correction involves a quite abstract formalism, which solves the final problem about the unwanted phase encountered in the derivation of the Schr\"{o}dinger equation, among other things.

\appendix

\section{Differential geometry}

I have included a basic introduction to differential geometry in case the reader is not familiar with the subject. There are many books on differential geometry; I recommend Schutz \cite{Schutz} on a first approach, followed by the more comprehensive Nakahara \cite{Nakahara}.

\paragraph{Manifolds.} A \emph{manifold} $M$ of dimension $n$ is, intuitively, a set of points which form a ``continuum.'' It is a generalization of the notion of a space. Some examples are the real line $\mathbb{R}$, or any open subset $(a,b)\subset \mathbb{R}$, the plane $\mathbb{R}^2$, the $n$-dimensional Euclidean space $\mathbb{R}^n$, the sphere $S^2$, the torus $T^2$, and $n$-dimensional versions thereof.\footnote{The 2-sphere $S^2$ always refers to the \emph{surface} of the sphere, rather than its interior, which is the $3$-ball $B^3$.} To the points $m$ of a manifold are assigned \emph{coordinates} $x(m)$, which are elements of $\mathbb{R}^n$, so that $x(m)$ is an $n$-tuple of real numbers. For example, the coordinates of a point $m$ on $S^2$ are typically denoted by $x(m) = (\theta,\phi)$. On a small enough \emph{patch} of any manifold, the patch ``looks'' like a Euclidean space (think about zooming in on the intersection of a line of latitude and a line of longitude on the sphere). However, one often encounters manifolds which require multiple patches in order to describe every point unambiguously. The circle is an example, which may be described by the two subsets $(0,2\pi), \; (-\pi, \pi)$ of $\mathbb{R}$.

Given a manifold $M$, the \emph{tangent space} $T_m M$ at a point $m \in M$ is a vector space spanned by the partial derivatives with respect to the coordinates $x$. This generalizes the notion of the tangent plane regarded as a plane resting against a surface at a point. That is, a vector $v$ tangent to $M$ at $m$ with coordinates $x$ may be expanded in a basis
\BE
v = v^a \frac{\del}{\del x^a}.
\EE
The tangent vectors naturally act on smooth functions $f \in C^\infty(M)$ as directional derivatives, $v[f] := v^a \del_a f$. The \emph{integral curves} of a vector field $v$ are the curves that follow the vector field over the manifold (thinking of $v$ as a flow of liquid, the integral curves are the curves traced out by objects floating along with the liquid), and are determined by the ODE system
\BE
\frac{\dd x^a}{\dd t} = v^a(x),
\EE
where $t$ parametrizes the curve. If the set of all integral curves of a vector field covers the entire manifold, the vector field is said to be \emph{complete}. The set of all tangent spaces $T_m M$ on $M$ is the \emph{tangent bundle} $TM$.

The directional derivative is also used to define the \emph{gradient} $\dd f$ by setting 
\BE
\dd f(v) := v[f] \quad \forall v \in TM.
\EE 
$\dd f$ is an example of a \emph{1-form}, which is the mathematician's name for a dual vector---a linear function of vectors which returns a value in $\mathbb{R}$ upon evaluation on a vector. The set of all 1-forms at a point $m$ is called the \emph{cotangent space} $T^*_m M$. Acting on the coordinates $x$, one finds $\dd x^a (v) = v^a$, so that for $v = \del_a$, one has the dual basis 1-form $\dd x^a$, and this basis satisfies a condition analogous to orthonormality, $\dd x^a (\del_b) = \delta^a_b$. A general 1-form $\theta$ can then be expanded in a basis as $\theta = \theta_a \dd x^a$, and the \emph{contraction} of $\theta$ with a vector $v$ is written variously as $v \intprod \theta = \theta(v) = v(\theta) = \theta_a v^a$. The set of all cotangent spaces $T^*_m M$ on $M$ is called the \emph{cotangent bundle} $T^*M$.

Another important kind of derivative is the \emph{Lie derivative} $\mcal{L}_X$ along a vector field $X$. Intuitively, it describes how an object changes along the flow of the vector field. On functions, it is defined by the directional derivative,
\BE
\mcal{L}_X f := X[f] \quad \in C^\infty(M).
\EE
On vectors, it is given by the Lie bracket, which is essentially just a commutator of differential operators,
\BE
\mcal{L}_X Y = [X,Y] \quad \in TM.
\EE
The action $\mcal{L}_X \theta \in T^* M$ on 1-forms is determined implicitly by the definitions above, together with the Leibniz axiom of derivations (the product rule),
\BE
(\mcal{L}_X \theta)(Y) := \mcal{L}_X (\theta(Y)) - \theta(\mcal{L}_X Y).
\EE
One then defines the action of $\mcal{L}$ on arbitrary tensors in the same way. Alternatively, one can use the notion of a pullback (discussed below) to define the Lie derivative of arbitrary tensors.

\paragraph{Tensors.} With the tangent and cotangent spaces, we can construct tensor products of them to produce the \emph{tensors} on $M$. For example, a type $(2,0)$ tensor is an object $T$ that can be expanded in the tensor product basis 
\BE
T = T^{ab} \del_a \otimes \del_b.
\EE
Since the basis vectors $\del_a$ act on 1-forms by $\theta(\del_a) = \theta_a$, the tensor $T$ naturally \emph{eats} two 1-forms to produce a function:
\BE
T(\theta, \eta) = T^{ab} (\del_a \otimes \del_b)(\theta,\eta) = T^{ab} \del_a(\theta) \del_b (\eta) = T^{ab} \theta_a \eta_b \quad \in C^\infty(M).
\EE
Similarly, a tensor of type $(1,1)$ acts on a vector \emph{and} a 1-form by
\BE
T(v,\theta) = T_a^b (\dd x^a \otimes \del_b)(v,\theta) = T_a^b v^a \theta_b \quad \in C^\infty(M).
\EE
One can go on to construct arbitrary tensor powers of the tangent and cotangent spaces, to form the tensor spaces $\mcal{T}^i_j (M)$ of type $(i,j)$. We can also form an anti-symmetric tensor product by setting, for 1-forms $\theta,\eta \in T^*M$,
\BE
\theta \wedge \eta := \theta \otimes \eta - \eta \otimes \theta.
\EE
This is called the \emph{wedge product}. The set of wedge products $\dd x^a \wedge \dd x^b$ span the space of antisymmetric type (0,2) tensors, denoted by $\Omega^2(M)$, which have the form
\BE
\omega = \omega_{ab} \dd x^a \wedge \dd x^b.
\EE
It follows that $\omega_{ab} = - \omega_{ba}$. An alternative definition is to say $\omega$ are the tensors such that $\omega(v,w) = -\omega(w,v)$ for all vectors $v, \; w$. One may further define the space $\Omega^p(M)$ of totally antisymmetric type $(0,p)$ tensors on $M$. The elements of $\Omega^p(M)$ are called $p$-\emph{forms}, and such tensors are referred to generally as \emph{differential forms}. They play a crucial role in differential geometry and physics. For example, the electromagnetic field strength tensor $F = F_{\mu\nu} \dd x^\mu \wedge \dd x^\nu$ on Minkowski space is a 2-form, and the Riemann curvature tensor $R$ may be regarded as a matrix-valued 2-form.

The gradient operator $\dd$ is extended to an \emph{exterior derivative} operator on arbitrary differential forms by taking an anti-symmetric derivative and increasing the rank of the form. For example, on the 2-form $\omega$ from above,
\BE
\dd \omega := \del_a \omega_{bc} \dd x^a \wedge \dd x^b \wedge \dd x^c \quad \in \Omega^3(M).
\EE
The exterior derivative allows for a simple expression of Stokes' theorem, as discussed below.

\paragraph{Integration.} 1-forms provide an elegant way of talking about line integrals. The \emph{line integral} of $\theta$ along a curve $\gamma$ whose tangent vector is $v$ is defined by
\BE
\int_\gamma \theta := \int_0^t v \intprod \theta \; \dd t = \int_0^t \theta_a v^a \dd t  \quad \Big(= \int_0^t \bo \theta \cdot \bo v \; \dd t\Big).
\EE
The last expression is the equivalent formula typically found in physics texts books, e.g., line integrals of a vector potential $\int \bo A \cdot \dd \bo r$, or in the definition of work $-\int \bo F \cdot \dd \bo r$. It is not common in differential geometry books, however. 

The $n$-forms on an $n$-dimensional manifold $M$ are called \emph{volume forms}, and have the form
\BE
\omega = f \dd x^1 \wedge \cdots \wedge \dd x^n, \quad f\in C^\infty(M).
\EE
They only have one independent component and therefore span a one-dimensional vector space. So long as $M$ is orientable, the $n$-form $\dd x^1 \wedge \cdots \wedge \dd x^n$ is called \emph{oriented} (this is a generalization of the notion that the cross product determines a unique direction at every point of $M$), and one defines volume integrals of functions $f$ by using ordinary integration on $\mathbb{R}^n$,
\BE
\int_M f \dd x^1 \wedge \cdots \wedge \dd x^n := \int_{x(M)} \!\!\!\! \!\!\! f(x) \dd^n x,
\EE
where $x(M)$ is the subset of $\mathbb{R}^n$ that the points of $m$ map to. (Technically, $M$ will in general require several coordinate patches, each having different coordinates, e.g., the sphere, which needs two patches in order to cover both poles, so that the right-hand side needs to be a sum over all of these patches.)

Once volume integration is defined, one can prove the general form of Stokes' theorem, which equates the integration of an exterior derivative to an integral over a boundary. Let $\mn\Sigma$ be any $d$-dimensional submanifold of $M$ with boundary $\del \mn\Sigma$, and let $\omega\in\Omega^{d-1}(M)$ be any $d$-form on $M$. Then Stokes' theorem states the equivalence
\BE
\int_{\mn\Sigma} \dd \omega = \int_{\del \mn\Sigma} \omega.
\EE

\paragraph{Pullbacks and pushforwards.} A \emph{diffeomorphism} is a mapping of points between manifolds. Let $\phi : M \rightarrow N$ be a diffeomorphism, which takes points of $M$ to points of $N$ in a smooth, differentiable fashion. Then points $p \in M$ are mapped to $\phi(p) \in N$. $\phi$ induces a pullback action on functions $f : N \rightarrow \mathbb{C}$ via composition,
\BE
(\phi^*f)(p) := (f \circ \phi)(p) =  f(\phi(p)).
\EE
As an example, consider the pullback of a function $f\in C^\infty(\mathbb{R}^3)$ to the 2-sphere $S^2$ of radius $r$. Let $\mn\Phi : S^2 \rightarrow \mathbb{R}^3$ be the map
\BE
(\theta,\phi) \longmapsto \mn\Phi(\theta,\phi) = (r\sin\theta\cos\phi, r\sin\theta \sin\phi, r\cos \theta) = \bo x(\theta,\phi).
\EE
Then the pullback just re-expresses the $\bo x$-dependence of $f$ in terms of $(\theta,\phi)$:
\BE
(\mn\Phi^* f)(\theta,\phi) = f(\bo x(\theta,\phi)).
\EE
This might seem like a triviality, that all we have done is rewrite the composition map. But in general, there are many nontrivial possibilities: sometimes $\phi(M) = N$, or $\dim M > \dim N$, and so on.

Since vector fields $X: C^\infty(M) \rightarrow \mathbb{C}$ are defined by their linear action on functions, we can define a pushforward map on vectors $X\in T M$ via
\BE
(\phi_* X)[f]|_{\phi(p)} := X[\phi^* f]|_p.
\EE
For example, with the map $\mn\Phi$ above, we may compute the pushforward of a vector on $S^2$ to one on $\mathbb{R}^3$. Let $X= X^\theta \del_\theta + X^\phi \del_\phi$ be a vector field on $S^2$, and let $f$ be a function on $\mathbb{R}^3$. The pushforward of $X$ to $T\mathbb{R}^3$ is determined by computing
\BE
X[f\circ\mn\Phi](\theta, \phi) = \big(X^\theta \del_\theta + X^\phi \del_\phi\big)[f(\bo  x(\theta,\phi))] = \Big( X^\theta \frac{\del x^i}{\del \theta} +X^\phi \frac{\del x^i}{\del \phi}\Big) \del_i  f(\bo x)|_{\bo x(\theta,\phi)}.
\EE
Although the vector field on the right-hand side depends explicitly on $(\theta,\phi)$, we may still regard it as a vector field on all the points $\bo x \in \mathbb{R}^3$ which coincide with points that are mapped to by $\mn\Phi$ from $S^2$. This is precisely the pushforward of $X$ by $\mn\Phi$, namely, $\mn\Phi_* X$. Lastly, the pushforward induces a pullback action on 1-forms $ \eta : TN \rightarrow \mathbb{C}$ as
\BE
(\phi^* \eta)(X)|_p := \eta(\phi_* X)|_{\phi(p)}.
\EE
For the map $\mn\Phi$, one can compute the pullback $\mn\Phi^* \eta$ of a 1-form in $T^* \mathbb{R}^3$ to find
\BE
\mn\Phi^*\eta  = \eta_i  \Big(\frac{\del x^i}{\del \theta} \dd \theta + \frac{\del x^i}{\del \phi} \dd \phi \Big) \quad \in T^* S^2.
\EE
One can check that $\dd \phi^* = \phi^* \dd$. The pullback also ``distributes'' into wedge products as
\BE
\phi^* (\eta_1 \wedge \cdots \wedge \eta_p) = \phi^* \eta_1 \wedge \cdots \wedge \phi^* \eta_p.
\EE
These maps play an important role in the definition of time evolution in both classical systems and GQ.

\section{Fiber bundles and connections} 

\paragraph{Fiber bundles.} A principle fiber bundle $P(M,G)$ is a manifold $P$ with a projection $\pi:P\rightarrow M$, where $M$ is called the \emph{base space}, and which looks \emph{locally} like the Cartesian product $M \times G$, where $G$ is a Lie group \cite{Nakahara}. The projection maps points $p$ in $P$ to points $\pi(p)$ in the base $M$. $P$ has coordinate maps called \emph{local trivilizations} $\phi_i$ over open subsets $U_i \subset P$. A point $p \in P$ has coordinates $\phi(p) = (m, g)$ where $m$ are the coordinates of $\pi(p) \in M$, and $g \in G$ is a coordinate on the fiber $\pi^{-1}(m)$ above $m$. 

A \emph{section} $s$ of $P$ is a smooth map $s: M \rightarrow P$, and is therefore denotable in a local trivialization by $\phi(s(m)) = (m, g(m))$.\footnote{I abuse notation by frequently omitting the map $\phi$ in what follows.} Roughly speaking, the $g(m)$ can be thought of as a function which has values in $G$, and which varies smoothly over $M$. Two sections $s_i(m) = (m, g_i(m))$ and $s_j(m) = (m, g_j(m))$ are related by \emph{transition functions}, which are $G$-valued objects $t_{ij}$ such that $g_j = t_{ij} g_i$. This is just the statement that for any two $g,h \in G$, there exists another $k \in G$ such that $g = kh$. The \emph{unit section} $\mathfrak{e}$ of a trivialization $\phi$ is defined by $\mathfrak{e}(m) = (m, e)$, where $e = \mrm{id}_G$. Thus, any other section can be written with respect to the unit section by $s(m) = g(m) \mathfrak{e}(m)$, where the action of $G$ on a section is the obvious $g(m,h) = (m, gh)$. In physics, the local trivializations appear as position-dependent phases or matrices $\me^{i\theta(x)} \in G$ that multiply things such as wave functions in an electromagnetic field, or complex fields with a local gauge symmetry in QFT.

\emph{Vector bundles} $\pi : E \rightarrow M$ are defined similarly, but $G$ is replaced by a vector space $V$, so that the local trivializations map $p\in E$ to $\phi(p) = (m, v)$, where $v\in V$, and $\pi(p) = m \in M$. Sections of $E$ are then vector fields $s(m) = (m, v(m))$, and the \emph{null section} is the zero vector field, $\mathfrak{0}(m) = (m, 0)$. A standard example is the tangent bundle $TM$ on an arbitrary manifold $M$. The fibers $\pi^{-1}(m)$ are simply the tangent spaces $T_m M$ at $m$, and sections are the vector fields $X$ on $M$.

Lastly, an \emph{associated vector bundle} $E(M;G \times_\rho V)$ is a combination of the prior two kinds of bundles; the fibers are locally products $G \times V$, and there is a projection $\pi_E: E \rightarrow M$. The group $G$ acts on $V$ by a representation $\rho$ as $\rho(g)v \in V$. One defines an equivalence relation $(g,v) \sim (hg, \rho(h)^{-1}v)$, and defines points of $E$ to be such equivalence classes. Elements of $E$ should therefore be denoted by $[(g,v)]$, where $(g,v)$ is merely a \emph{representative} of that class. The physical reason for this seemingly odd construction is that, in any theory with a local symmetry, the functions $\psi(x)$ and $g(x)\psi(x)$ are physically equivalent (think local phases of a wave function in an electromagnetic field, or local changes of basis for a complex field). Choosing a particular representative $(g,v)$ in a calculation is called \emph{fixing the gauge}. Sections are denoted by $s(m) = [(g(m), v(m))]$, but we will be sloppy about explicitly including the equivalence class notation throughout this paper. The space of sections on a vector bundle is denoted by $\Gamma(E)$.

\paragraph{Connections.} A connection $\nabla$ on an associated vector bundle can be defined axiomatically. It is a map $\nabla : \Gamma(E) \rightarrow \Gamma(E) \otimes \Omega^1(M)$, $s \mapsto \nabla s$, satisfying several properties:
\BE
\nabla(as+bs') = a\nabla s + b\nabla s', \quad \nabla (fs) = \dd f \otimes s + f \nabla s, \quad \nabla_{fX+gY} = f\nabla_X + g\nabla_Y,
\EE
for constants $a,b,$ functions $f,g\in C^\infty(M)$, and vectors $X,Y \in TM$. Its 1-form character is described by $X \intprod \nabla s = \nabla_X s$ for $X\in TM$, that is, the covariant derivative of $s$ along $X$. For general $p$-forms $\alpha \in \Omega^p (M)$, $\alpha s := \alpha \otimes s$ is a $V$-valued $p$-form (which is a section of the product bundle $\Omega^p(M) \times E$). On a vector bundle with fibers $V$, one then defines the \emph{connection 1-form} $\Theta \in \Omega^1(M)\otimes \mathfrak{g}$ by its action on an orthonormal basis $\{ \me_a \}$ of $V$ by\footnote{Alternatively, we can define (a l\'{a} Ehresmann) the $\mathfrak{g}$-valued connection 1-form $\tilde \Theta$ on the cotangent space $T^* P$ of the principal bundle which divides vertical and horizontal vector bundles on $P$. One typically deals with the pullback $s^* \tilde \Theta = \Theta$ by a section $s$ to the manifold in physics applications.}
\BE
\nabla_X \me_a:= \Theta(X) \me_a = \Theta(X)^b_a \me _b
\EE
where $a = 1,...,\dim V$, so that the $a,b$ are indices in the representation of the Lie algebra $\mathfrak{g}$ on $V$. We can also write the connection as $\Theta = \Theta_\mu \dd x^\mu$, where $\Theta_\mu = \del_\mu \intprod \Theta$ is the $\mathfrak{g}$-valued matrix $\Theta_\mu = [\Theta^a_{\mu b}]$, and $\del_\mu$ is a basis vector of $TM$. For example, if the group $G$ is the set of coordinate transformations $y = f(x)$ with representation $\del y^\mu / \del x^\nu$ on $TM$, we recover the familiar formula $\nabla_\mu \del_\nu = \Gamma^\lambda_{\mu\nu} \del_\lambda$ from general relativity, and if $G$ is a Lie group acting on complex scalar fields $\Phi \in \mathbb{C}^N$, then $\nabla_\mu \Phi = \del_\mu \Phi + A_\mu \Phi$ where $A_\mu$ is the $N\times N$ matrix gauge field (but we have to use the Leibniz axiom for $\nabla_X (f s)$).

The \emph{curvature} on the bundle is then given by the pullback of a covariant exterior derivative $\Omega := s^* D_\Theta \Theta$ \cite{Nakahara}, or equivalently by its action on sections of $E$,
\BE
\Omega(X,Y) s:= \big([\nabla_X, \nabla_Y] - \nabla_{[X,Y]}\big) s.
\EE
Since $\Omega(X,Y)$ naturally acts on sections to produce another section, which in turn eat dual sections, all linearly, $\Omega$ is sometimes regarded as a type $(1,3)$ tensor in $\mathcal{T}^1_3(M)$ (which explains, for example, the notation in physics literature, $R^\mu_{\; \nu \alpha \beta}$, for the Riemann tensor, where $E$ is the tangent bundle, and sections are vector fields). Alternatively, the exterior derivative definition implies that $\Omega \in \Omega^2(M) \otimes \mathfrak{g}$, so $\Omega$ is a Lie algebra-valued 2-form on  $M$.

\paragraph{Line bundles.} The formalism of GQ uses extensively the structure of a \emph{line bundle}, which is an associated bundle over $M$ with 1-dimensional fibers. For fibers $G = U(1)$, the Lie algebra is one dimensional, $\mathfrak{g} = i\mathbb{R} \cong \mathbb{R}$. We take as our vector space $V = \mathbb{C}$ for simiplicity, which is the space our wave functions live in. The connection 1-form is then determined by its action on the unit section $\mathfrak{u}(m) = \phi^{-1}(m,1)$, $1\in\mathbb{C}$, by
\BE
\nabla_X \mathfrak{u} := -i\Theta(X) \mathfrak{u}.
\EE
The $-i$ is conventionally put there with the foresight that $\Theta$ will be real if the connection is compatible with a Hermitian structure $(\cdot,\cdot)$ (basically a point-wise inner product), that is,
\BE
\nabla_X (s,s') = (\nabla_X s, s') + (s, \nabla_X s').
\EE
This is the analog of metric compatibility $\nabla_X \mrm{g} = 0$ on Riemannian manifolds, but here the vectors are not tangent vectors on $M$, instead they are complex vectors, such as wave functions. Sections of $P(M,G)$ are $U(1)$-fields, $\me^{i\theta(m)}$. Sections of the associated bundle are complex vectors, which we conventionally write in terms of the unit section as $s_\psi(m) = \psi(m) \mathfrak{u}(m)$, $\psi\in\mathbb{C}$. Covariant derivatives of general sections are then given by 
\BE
\nabla_X s_\psi = X[\psi] \mathfrak{u} - i \psi \Theta(X) \mathfrak{u} = X \intprod (\dd \psi - i \Theta \psi) \mathfrak{u}.
\EE
The curvature is then given by
\BE
-i\Omega(X,Y)s = \big( [\nabla_X, \nabla_Y] - \nabla_{[X,Y]} \big) s.
\EE

\section{K\"{a}hler manifolds} 

Let $(M,J)$ be a complex manifold with complex dimension $\dim_\mathbb{C} M = n$ and complex structure $J$. The $2n$ real coordinates are denoted by $\{q^a,p_a\}$, $a=1,...,n.$ The structure $J$ is defined by
\BE
J \frac{\del}{\del q^a} = \frac{\del}{\del p_a}, \quad J \frac{\del}{\del p_a} = - \frac{\del}{\del q^a},
\EE
so that $J^2 = - \mathbb{I}$. There exists locally a \emph{holomorphic} coordinate system
\BE
z^a = p_a + i q^a, \quad \bar z^a = p_a - i q^a, \quad a=1,\dots,n.
\EE  
$J$ divides the (complexified) tangent space into a direct sum $T_m M = T^+_m M \oplus T^-_m M$ at $m\in M$, distinguished by the eigenvalues $\pm i$ of $J$:
\BE
T^\pm_m M = \{ Z \in T_m M | \; J Z = \pm i Z  \}
\EE
The bases for each component $T_m^\pm M$ are given explicitly by
\BE
\frac{\del}{\del z^a} = \frac{1}{2} \Big( \frac{\del}{\del p_a} - i \frac{\del}{\del q^a} \Big), \quad \frac{\del}{\del \bar z^a} = \frac{1}{2} \Big( \frac{\del}{\del p_a} + i \frac{\del}{\del q^a} \Big),
\EE
where the 1/2 is conventional. We sometimes abbreviate these vectors by $\del_a, \; \del_{\bar a}$. The corresponding dual basis is given by
\BE
\dd z^a = \dd p_a + i \dd q^a, \quad \dd \bar z^a = \dd p_a - i \dd q^a.
\EE

If $\mrm{g}$ is a metric on $M$ that is compatible with $J$, in the sense that $\mrm{g}(JX,JY) = \mrm{g}(X,Y) \; \forall X,Y \in TM$, then there exists a rank-2 tensor $\Omega$ such that
\BE
\Omega(X,Y) := g(JX,Y)
\EE
defines a 2-form, and $\Omega$ is called the K\"{a}hler form \cite{Nakahara}. The pair $(M,\mrm{g},J)$ is called a \emph{Hermitian manifold}. The Dolbeaux operators $\del, \; \bar \del$ are the holomorphic, antiholomorphic exterior derivatives such that $\dd = \del + \bar \del$. If $\dd \Omega = 0$, then locally, 
\BE
\Omega = i\del \bar \del \mcal{K},
\EE
and the real function $\mcal{K}$ is called the K\"{a}hler scalar. Such manifolds $(M,g,J,\mcal{K})$ are called \emph{K\"{a}hler manifolds}. In particular, $M = \mathbb{C}^n$ is K\"{a}hler, and the 2-form and scalar are
\BE
\Omega = \frac{i}{2} \dd z^a \wedge \dd \bar z^a, \quad \mcal{K} = \frac{1}{2} z^a \bar z^a.
\EE
Thus, any K\"{a}hler manifold will have \emph{local} coordinates for which the K\"{a}hler structure is given this way. The two-form determines a symplectic structure given in real coordinates by $\omega = \dd p_a \wedge \dd q^a$. 


\end{document}